\documentclass[twocolumn]{aastex62}

\usepackage{tikz}
\usepackage{tikz-qtree}

\accepted{2019 November 1}
\submitjournal{AJ}

\shorttitle{The Feasibility of Directly Imaging Nearby Cold Jovian Planets with MIRI/JWST}
\shortauthors{Brande et al.}

\begin{document}

\title{The Feasibility of Directly Imaging Nearby Cold Jovian Planets with MIRI/JWST}

\correspondingauthor{Jonathan Brande}
\email{jonathan.brande@nasa.gov}

\author[0000-0002-2072-6541]{Jonathan Brande}
\altaffiliation{Sellers Exoplanet Environments Collaboration}
\affiliation{NASA Goddard Space Flight Center, 8800 Greenbelt Rd, Greenbelt, MD 20771}
\affiliation{University of Maryland, College Park, MD 20742}

\author[0000-0001-7139-2724]{Thomas Barclay}
\altaffiliation{Sellers Exoplanet Environments Collaboration}
\affiliation{NASA Goddard Space Flight Center, 8800 Greenbelt Rd, Greenbelt, MD 20771}
\affiliation{University of Maryland, Baltimore County, 1000 Hilltop Cir, Baltimore, MD 21250}

\author[0000-0001-5347-7062]{Joshua E. Schlieder}
\altaffiliation{Sellers Exoplanet Environments Collaboration}
\affiliation{NASA Goddard Space Flight Center, 8800 Greenbelt Rd, Greenbelt, MD 20771}

\author{Eric D. Lopez}
\altaffiliation{Sellers Exoplanet Environments Collaboration}
\affiliation{NASA Goddard Space Flight Center, 8800 Greenbelt Rd, Greenbelt, MD 20771}

\author{Elisa V. Quintana}
\altaffiliation{Sellers Exoplanet Environments Collaboration}
\affiliation{NASA Goddard Space Flight Center, 8800 Greenbelt Rd, Greenbelt, MD 20771}

\begin{abstract}
The upcoming launch of the James Webb Space Telescope (JWST) will dramatically increase our understanding of exoplanets, particularly through direct imaging. Microlensing and radial velocity surveys indicate that some M-dwarfs host long period giant planets. Some of these planets will likely be just a few parsecs away and a few AU from their host stars, a parameter space that cannot be probed by existing high-contrast imagers. We studied whether the coronagraphs on the Mid-Infrared Instrument on JWST can detect Jovian-type planets around nearby M-dwarfs. For a sample of 27 very nearby M-dwarfs, we simulated a sample of Saturn--Jupiter-mass planets with three atmospheric configurations, three orbital separations, observed in three different filters. We found that the f1550c $15.5\mu$m filter is best suited for detecting Jupiter-like planets. Jupiter-like planets with patchy cloud cover, 2 AU from their star, are detectable at $15.5\mu$m around 14 stars in our sample, while Jupiters with clearer atmospheres are detectable around all stars in the sample. Saturns were most detectable at 10.65 and $11.4\mu$m (f1065c and f1140c filters), but only with cloud-free atmospheres and within 3 pc (6 stars). Surveying all 27 stars would take $<170$ hours of JWST integration time, or just a few hours for a shorter survey of the most favorable targets. There is one potentially detectable known planet in our sample -- GJ~832~b. Observations aimed at detecting this planet should occur in 2024--2026, when the planet is maximally separated from the star.

\end{abstract}


\section{Introduction} 
\label{sec:intro}
Direct imaging is a powerful technique for exoplanet detection and characterization. Planets with wide orbital separations are difficult to detect with time-series methods such as transit photometry and radial velocity measurements. However, direct imaging can detect planets with sufficient brightness and orbital separations more quickly than other methods, and, depending on the capabilities of the instrument, over more of the planet's orbit. Unfortunately, imaging planets requires very high-contrast imaging capabilities, which make direct imaging very challenging for all but the most favorable targets. Ground-based, adaptive optics-equipped, large-diameter telescopes and high-contrast coronagraphs are the main drivers of direct imaging exoplanet detection, accounting for most of the directly imaged exoplanets currently known. 
These include notable discoveries such as four planets orbiting HR~8799 \citep{marois2008, marois2010}, Beta~Pictoris~b \citep{lagrange2009}, HD~95086~b \citep{rameau2013a,rameau2013b}, HIP~65426~b \citep{chauvin2017}, 51~Eridani~b \citep{macintosh2015}, and the first directly imaged protoplanet, PDS~70~b \citep{keppler2018}. 
Space-based direct imaging has also been successfully demonstrated: planets have been directly imaged with the Spitzer Space Telescope \citep[HN~Pegasi~b, FU~Tauri~b, WD~086-661~b; ][]{luhman2007,luhman2009,luhman2011} and the Hubble Space Telescope \citep[CXHR~73~b, Fomalhaut~b, 2MASS J04414489+2301513 b; ][]{luhman2006,kalas2008,todorov2010}. As of March 2019, where mass and semimajor axis data are available in the Exoplanet Archive \citep{ExoplanetArchive}, all directly imaged exoplanets are more massive then Jupiter, almost all are all on orbits wider than 10 AU, and almost all are young planets that still retain much of the internal heat from their formation. 

Current ground-based high-contrast imaging instruments have poor sensitivity at wavelengths needed to probe older, cooler, fainter planets, and as such are limited to detecting planetary mass companions around younger stars where the planets are warmer and brighter at shorter wavelengths. These younger host stars are typically found further away, and when combined with the sensitivity limits, generally lead to detections of wider orbit, more massive planets. Existing space-based imagers are similarly insensitive to cooler targets, and are more limited to the red-optical/near-infrared bands. As such, upcoming space-based mid-infrared direct imaging has the potential to focus on older and closer stars and allow the detection of cooler, fainter, lower-mass planetary companions at smaller projected separations.

The upcoming James Webb Space Telescope (JWST) will be highly capable of high-contrast imaging. Three instruments on the spacecraft are imaging-capable: the Near-Infrared Camera (NIRCam), the Near-Infrared Imager and Slitless Spectrograph (NIRISS), and the Mid-Infrared Instrument (MIRI). NIRCam and MIRI are equipped with coronagraphs, while NIRISS is capable of high-contrast imaging through its Aperture Masking Interferometry imaging mode.

\subsection{Planned JWST Observations}
\citet{beichman2010} studied the general capabilities of JWST's imaging modes compared with ground-based facilities, and found that the MIRI instrument on JWST would be capable of detecting planets with masses as small as 0.8 $M_{J}$ and with semimajor axes greater than 4 AU. However, these estimates are worth revisiting because they rely on predicted JWST instrument parameters of the time. More recent work on estimating the performance of the instruments and re-evaluated exoplanetary formation and evolution models allows us to revise estimates of JWST's capabilities. 

Several JWST Early Release Science (ERS) and Guaranteed Time Observation (GTO) programs focus on exoplanet direct imaging.
Beichman et al. will perform a three-part study as part of their GTO program to coronagraphically image exoplanets and debris disks. They will investigate three nearby well-known debris disks with MIRI and simultaneously search for planets with NIRCam\footnote{\url{https://www.stsci.edu/jwst/phase2-public/1193.pdf}}, further characterize the HR 8799 system and also search for new planets\footnote{\url{https://www.stsci.edu/jwst/phase2-public/1194.pdf}}, and characterize the known planet HD 95086 b using NIRCam, and also search for undiscovered companions around the star\footnote{\url{https://www.stsci.edu/jwst/phase2-public/1195.pdf}}. Proposal 1194 will push the capabilities of the MIRI coronagraphs by attempting to image HR 8799 e at a separation of only about 300 mas ($\sim14.5$ AU). Ressler et al.\footnote{\url{https://www.stsci.edu/jwst/phase2-public/1241.pdf}} focus on recovering previously known exoplanets, 51 Eri b, HR 8799 e, and $\kappa$ And b. Partially motivating our own work, Schlieder et al.\footnote{\url{https://www.stsci.edu/jwst/phase2-public/1184.pdf}} proposed a short NIRCam survey of young nearby M-dwarfs to search for young giant planets on wide orbits ($\gtrsim$ 10 AU) down to the Saturn-mass regime and potentially down to the Neptune-mass regime for the most favorable targets \citep{schlieder2015}. In general, these programs focus on recovering previously discovered exoplanets, although several of them also intend to search for undiscovered exoplanets, including the M dwarf program described above.

These programs all perform observations that are impractical or impossible from current ground- and space-based high-contrast imaging facilities. However, they do not probe every part of the mass/separation space around nearby M-dwarfs. It would be more challenging, yet still likely feasible, to detect planets in the lower-mass and closer-orbiting areas of the detectability window theorized by \citet{beichman2010}. In this paper we perform a systematic study determining whether it is feasible for JWST to image cold planets around some of the nearest M-dwarfs to the Solar System using the MIRI instrument. For the purposes of this feasibility study, we do not simulate an actual direct imaging survey. We do not inject planets around our host stars using measured occurrence rates, and neither do we randomize our orbital parameters. We instead opt to sample along a range of parameters that include stellar host distance and discrete projected separations, planet masses, and atmospheric models to investigate the limits of the current JWST sensitivity estimates given favorable exoplanetary targets. Given the idealized assumptions made by the current versions of the JWST planning tools (see Section \ref{sec:simulating_planets}), this feasibility study provides an upper limit on coronagraphic detections of the modeled planets around some of the nearest M-dwarfs.

\subsection{Demographics of giant exoplanets around low-mass stars}

\begin{figure*}[ht!]
\plotone{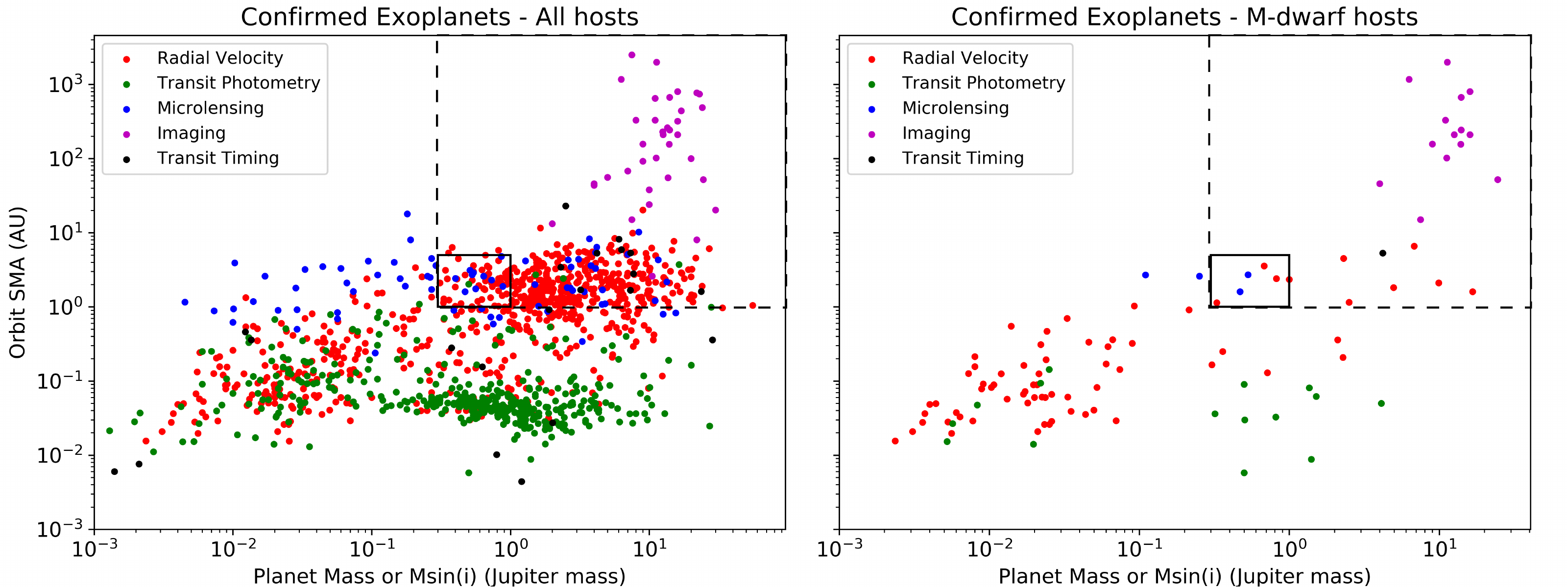}
\caption{Exoplanet semimajor axis vs. planet mass ($M_{J}$) for all stellar hosts (left panel) and M-dwarf hosts(right-panel). The parameter space that we study in this paper is bracketed with solid black rectangles, dashed black rectangles indicate the broader sensitivity space of MIRI coronagraphic imaging. Our focus is on planets with masses between $0.3 M_{J}$ -- $1 M_{J}$ (Saturn mass to Jupiter mass) at 1 - 5 AU separations. There are a small number of known M dwarf planets in this range, all detected by the radial velocity or microlensing techniques. JWST MIRI opens a new parameter space for direct imaging.}
\label{figure:planets}
\end{figure*}

While there is strong evidence that giant planets are rarer than smaller planets \citep{Burke2015}, and are less common around M-dwarfs than FGK-dwarfs \citep{Mulders2015}, many giant planets around M-dwarfs have been found. In Figure~\ref{figure:planets} we show known exoplanets in the left panel and planets orbiting cool stars in the right panel, demonstrating the sample of giant planets around cool stars. Some notable Jupiter-sized planets orbiting M-dwarfs are Kepler-45 b \citep{johnson2012}, HATS-6 b \citep{hartman2015}, NGTS-1 b \citep{bayliss2018}, and HATS-71 b \citep{bakos2018}. Radial velocity observations have detected over a dozen giant planets around M-dwarfs \citep{butler2006,johnson2007,bailey2009,apps2010,howard2010,johnson2010,haghighipour2010,forveille2011,johnson2012,robertson2013,hartman2015,sahlmann2016,bryan2016,trifonov2018,bayliss2018,bakos2018}. Microlensing observations are typically sensitive to planets orbiting between 1--10 AU from their host stars, and due to the statistical dominance of M-dwarfs in the galaxy, most microlensing-detected planets orbit low-mass stars \citep{zhu2017}. Early results from microlensing surveys, such as \citet{cassan2012}, found a giant planet occurrence rate of $17^{+6}_{-9} \%$ for planets with masses $0.3-10 M_{J}$ at distances between 0.5 and 10 AU. More recent microlensing results further reveal that giant planets are relatively common, although the planet mass function is steep \citep{suzuki2016, shvartzvald2016, mroz2017}, with many more Neptune-mass planets than Jupiter-mass planets, and tentative signs of a break between the ice-giant ($\sim$15 M$_{\oplus}$) and super-Earth ($\sim$5 M$_{\oplus}$) mass regimes \citep{suzuki2016, udalski2018}.

Combining RV and direct imaging observations, \citet{montet2014} estimated that $6.5\pm3.0$\% of M-dwarfs host a $1-13 M_{J}$ planet between 0 and 20 AU, while \citet{clanton2016} and 
\citet{meyer2018} find results consistent with previous microlensing studies, and that giant planets are more common at wider separations around low-mass stars. These results paint a similar picture for M-dwarfs as for FGK-dwarfs, where $6.2^{+2.8}_{-1.6}\%$ of all types of stars host giant planets between 3-7 AU \citep{wittenmyer2016}. Beyond about 10 AU, planets are scarce: \citet{lannier2016} found that the planetary mass companion frequency was $2.3^{+2.9}_{-0.7}\%$ between 8 and 400 AU, which is consistent with the findings from \citet{bowler2016} that $<3.9\%$ of M-stars host a $5-13 M_{J}$ planet between 30-300 AU.
While giant planets are relatively uncommon around M-dwarfs, they do exist a few AU from their stars at an occurrence rate in the range of 5--20\%, which motivates us to search for companions, and a sufficiently sensitive imaging survey of the nearest M dwarfs to Earth has the potential to yield a handful of such planets. 

\subsection{The MIRI Instrument and Direct Imaging}

Our primary interest is detecting field-aged planets around the nearest stars. These planets are likely to have temperatures of 100--300 K, similar to the giant planets in the Solar System, corresponding to peak blackbody wavelengths of 10--29 $\mu$m. The MIRI instrument on JWST covers 4.9 to 28.8 $\mu$m \citep{rieke2015}, which is well suited to our parameter space of interest. MIRI has four coronagraphs: three are 4-Quadrant Phase Masks optimized for use at 10.65 $\mu$m, 11.4 $\mu$m, and 15.5 $\mu$m, and have small inner working angles (IWAs) of 0.33$''$, 0.36$''$, and 0.49$''$ (equal to the 1 $\lambda / D$ limit) respectively. The fourth is a traditional Lyot spot for use at 23 $\mu$m which we do not consider due to its hard stop and large inner working angle of $2.16''$ \citep{boccaletti2015}.

For the three 4-Quadrant Phase Mask coronagraphs, the IWA is defined as the as the 50\% transmission radius. Our simulations (see Section \ref{sec:results}) reveal that for particularly suitable exoplanets at projected separations inside the IWA, the signal-to-noise ratio could indicate a photometric detection although the planet will not be spatially separated from the host star. Further discussion regarding the feasibility of sub-IWA planet detections is provided in Appendix \ref{appendix:sub_iwa}.

Table \ref{table:limit_dists} outlines the performance of the three coronagraphs. For the 10.65 $\mu$m coronagraph with an IWA of 0.33$''$, a planet orbiting at 1 AU falls within the IWA at host star distances beyond 3.03 pc. However, even planets 5 AU from their host stars would be within the IWA of the instrument if they are further than 15 pc.

\begin{figure}[ht!]
\plotone{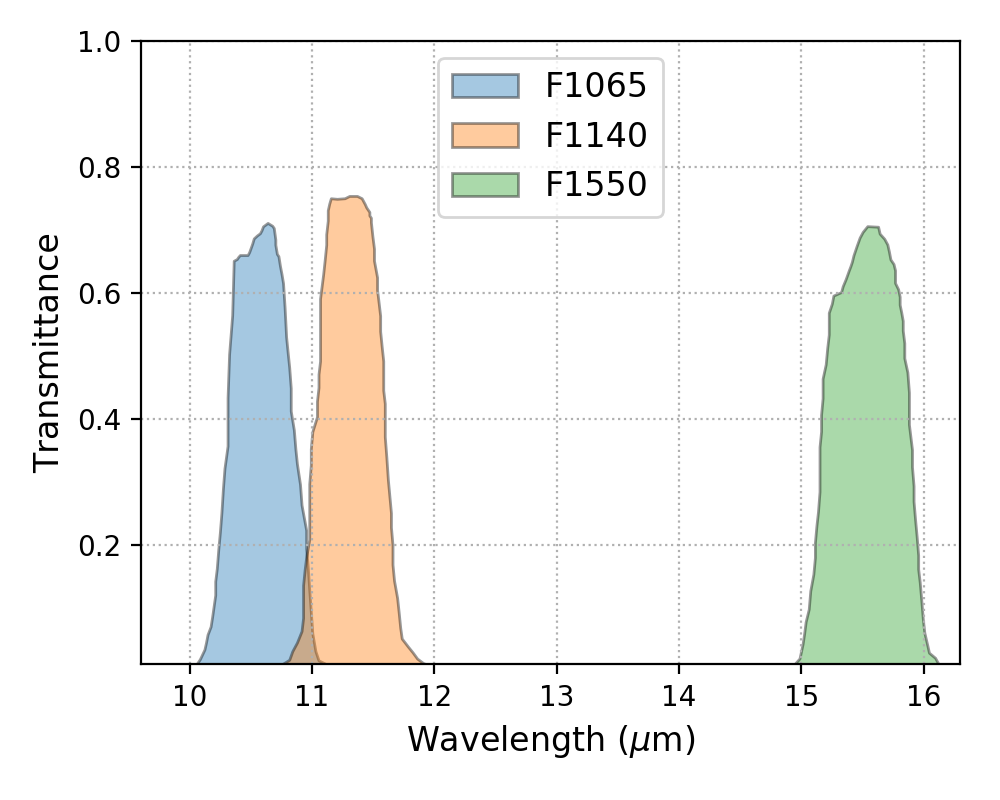}
\epsscale{0.10}
\caption{The optical transmission curves for the three MIRI coronographs that we use in this investigation. The transmission curves are adapted from \citet{Bouchet2015}. These three 4-quadrant phase mask coronagraphs have inner working angles of less than 0.5$''$.}
\label{figure:tran_curves}
\end{figure}

\begin{deluxetable}{lcrrr}
\tablecaption{Limiting target distances for each MIRI 4-Quadrant Phase Mask coronagraph. The system distances indicate where the projected separation is equivalent to the IWA for each filter.}
\label{table:limit_dists}
\tablehead{\colhead{Filter} & \colhead{IWA} & \colhead{1 AU} & \colhead{2 AU} & \colhead{5 AU}} 
\startdata
    f1065c & $0.33''$ & 3.03 pc & 6.06 pc & 15.15 pc \\
    f1140c & $0.36''$ & 2.7 pc & 5.5 pc & 13.8 pc \\
    f1550c & $0.49''$ & 2.04 pc & 4.08 pc & 10.2 pc \\
\enddata
\end{deluxetable}

\section{Simulating planets orbiting nearby cool stars}
\label{sec:simulating_planets}

Direct imaging measurements require the detection of faint planetary sources around much brighter stars. While previous studies typically concentrate on planets that still retain heat from their formation, this goal of this investigation is to concentrate on cold planets. This naturally leads us to nearby and intrinsically dim stars where projected separation is widest and the planet-to-star brightness ratio is most favorable.

We selected a population of stars from the RECONS list of the 100 nearest star systems to Earth\footnote{The RECONS 100 closest stars list was retrieved from \url{http://www.recons.org}}. Of the nearest M-dwarf stellar systems, we excluded those in tight binaries due to the observational and dynamical complications that would produce. Proxima Centauri was included despite its presence as the third component of the Alpha Centauri system because of its extremely wide separation from Alpha Centauri A and B. Although Proxima Centauri is not known to have any giant exoplanets, and is unlikely to have any undiscovered wide-orbit giant exoplanets, it makes for a good test calculation. We selected 27 stars in the RECONS list that fit our criteria. These stars and their properties are listed in Table~\ref{table:stellar_sample}.

To use these 27 stars as simulated exoplanet hosts, we needed to compile their stellar parameters.
The parameters were found, where possible, by using default parameters on SIMBAD \citep{wenger2000} and NASA's Exoplanet Archive \citep{ExoplanetArchive}. When parameters were not found in SIMBAD or the Exoplanet Archive, they were found from individual references in the literature, or estimated from comparisons to other similar stars in our sample. Although not every parameter was found, comparisons with similar systems show that the calculations from those systems are similar enough to not invalidate the general results. Table \ref{table:stellar_sample} shows the full set of stellar parameters we used to create our sample of targets.

\begin{deluxetable*}{lccrrrrrr}
\tablecaption{Stellar targets used as simulated exoplanet hosts}
\tablehead{\colhead{Name} & \colhead{Type} & \colhead{Temp} & \colhead{Mag} & \colhead{Mag} & \colhead{Mass} & \colhead{Radius} & \colhead{Distance} & \colhead{References} \\
\colhead{} & \colhead{} & \colhead{(K)} & \colhead{(V)} & \colhead{(J)} & \colhead{($M_{Sun}$)} & \colhead{($R_{Sun}$)} & \colhead{(pc)} & \colhead{}}
\startdata
Proxima Cen & M5.5V & 3050 & 11.13 & 5.357 & 0.12 & 0.141 & 1.29 & 1,2 \\ \hline
    GJ 699 & M4.0V & 3134 & 9.511 & 5.244 & 0.144 & 0.196 & 1.82 & 1,4,5,6 \\ \hline
    Wolf 359 & M5.5V & 2800 & 13.53 & 7.085 & 0.09 & 0.16 & 2.0 & 1,3,7  \\ \hline 
    GJ 411 & M2.0V & 3563 & 7.52 & 4.2 & 0.46 & 0.393 & 2.56 & 1,3,5  \\ \hline 
    Ross 154 & M3.5V & 3240 & 10.495 & 6.222 & 0.17 & 0.24 & 2.97 & 1,3,8  \\ \hline 
    Ross 248 & M5.0V & 3280 & 12.28 & 6.884 & 0.12 & 0.16 & 3.0 & 1,3,8  \\ \hline 
    GJ 887 & M2.0V & 3588 & 7.34 & 4.34 & 0.5 & 0.46 & 3.29 & 1,3,5  \\ \hline 
    Ross 128 & M4.0V & 3350 & 11.153 & 6.505 & 0.17 & 0.197 & 3.36 & 1,2  \\ \hline 
    DX Cnc & M6.5V & 3376 & 14.81 & 8.235 & 0.09 & 0.11 & 3.58 & 1,3,9  \\ \hline 
    GJ 1061 & M5.5V & 3000\tablenotemark{a} & 13.07 & 7.523 & 0.113 & 0.1\tablenotemark{a} & 3.67 & 1,3  \\ \hline 
    YZ Ceti & M4.0V & 3056 & 12.074 & 7.258 & 0.13 & 0.168 & 3.69 & 1,2,3  \\ \hline 
    Luyten's & M3.5V & 3317 & 9.872 & 5.714 & 0.26 & 0.35 & 3.8 & 1,3,10  \\ \hline 
    GAT 1370 & M7.0V & 3505 & 15.4 & 8.394 & 0.08 & 0.13 & 3.83 & 1,11,12  \\ \hline 
    Kapteyn's & M2.0V & 3722 & 8.853 & 5.821 & 0.27 & 0.29 & 3.91 & 1,2  \\ \hline 
    GJ 628 & M3.0V & 3272 & 10.072 & 5.95 & 0.294 & 0.307 & 4.29 & 1,2  \\ \hline 
    GJ 1 & M2.0V & 3400 & 8.562 & 5.328 & 0.48 & 0.48 & 4.34 & 1,3,13  \\ \hline 
    LHS 292 & M6.5V & 2772 & 15.784 & 8.857 & 0.08 & 0.1\tablenotemark{a} & 4.5 & 1,3  \\ \hline 
    GJ 674 & M3.0V & 3257 & 12.177 & 5.711 & 0.36 & 0.42 & 4.54 & 1,3,13  \\ \hline 
    GJ 687 & M3.0V & 3439 & 9.15 & 5.335 & 0.401 & 0.492 & 4.55 & 1,14  \\ \hline 
    GJ 876 & M3.5V & 3444 & 10.192 & 5.834 & 0.37 & 0.38 & 4.69 & 1,15  \\ \hline 
    LHS 288 & M4.0V & 3000 & 13.92 & 8.492 & 0.11 & 0.1\tablenotemark{a} & 4.79 & 1,3,16  \\ \hline 
    GJ 1002 & M5.5V & 3292 & 13.837 & 8.323 & 0.11 & 0.1\tablenotemark{a} & 4.83 & 1,3,11  \\ \hline 
    GJ 832 & M2.0V & 3419 & 8.672 & 5.349 & 0.45 & 0.48 & 4.95 & 1,2,8  \\ \hline 
    GJ 388 & M4.0V & 3380 & 9.52 & 5.449 & 0.39 & 0.39 & 4.96 & 1,3,17  \\ \hline 
    VB 10 & M8.0V & 2700 & 17.3 & 9.908 & 0.07 & 0.102 & 5.0 & 1,3,18  \\ \hline 
    LP 944-20 & M9.5V & 2040 & 18.69 & 10.725 & 0.07 & 0.1\tablenotemark{a} & 5.0 & 1,3  \\ \hline 
    VB 8 & M7.0V & 2700 & 16.916 & 9.776 & 0.08 & 0.1\tablenotemark{a} & 6.4 & 1,3  \\ \hline 
\enddata
\label{table:stellar_sample}
\tablerefs{(1) SIMBAD \citep{wenger2000}, (2) NASA Exoplanet Archive \citep{ExoplanetArchive}, (3) Research Consortium on Nearby Stars \url{http://www.recons.org}, (4) \citet{bobylev2010}, (5) \citet{demory2009}, (6) \citet{dawson2004}, (7) \citet{doyle1990}, (8) \citet{johnson1983}, (9) \citet{morin2010}, (10) \citet{lacy1977}, (11) \citet{davison2015}, (12) \citet{dieterich2014}, (13) \citet{pasinetti2001}, (14) \citet{berger2006}, (15) \citet{vonbraun2014}, (16) \citet{henry2006}, (17) \citet{reiners2009}, (18) \citet{linsky1995}}
\tablenotetext{a}{assumed values based on available stellar properties}
\end{deluxetable*}

\begin{figure}[ht!]
    \centering
    \plotone{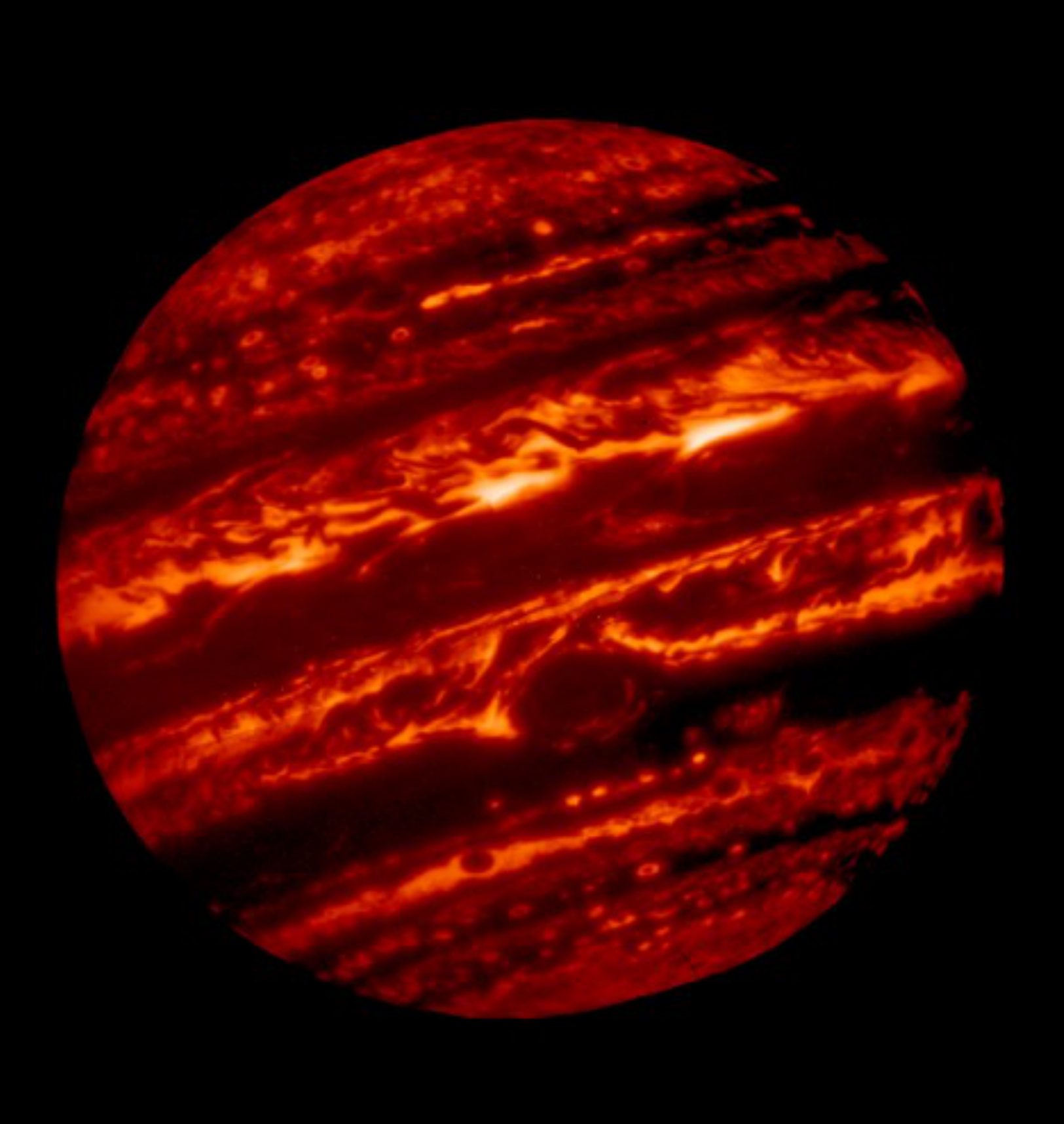}
    \caption{Jupiter as observed in 4.8 $\mu$m by the Near-Infrared Imager at Gemini North.
    In the infrared, the patchy nature of the clouds is striking. The thermal emission is absorbed by the cloud covered regions and escapes to be observed where there are gaps. Credit: Gemini Observatory/AURA/NSF/UC Berkeley.
    }
    \label{fig:jupiter_gemini}
\end{figure}

\begin{figure*}[ht!]
\plotone{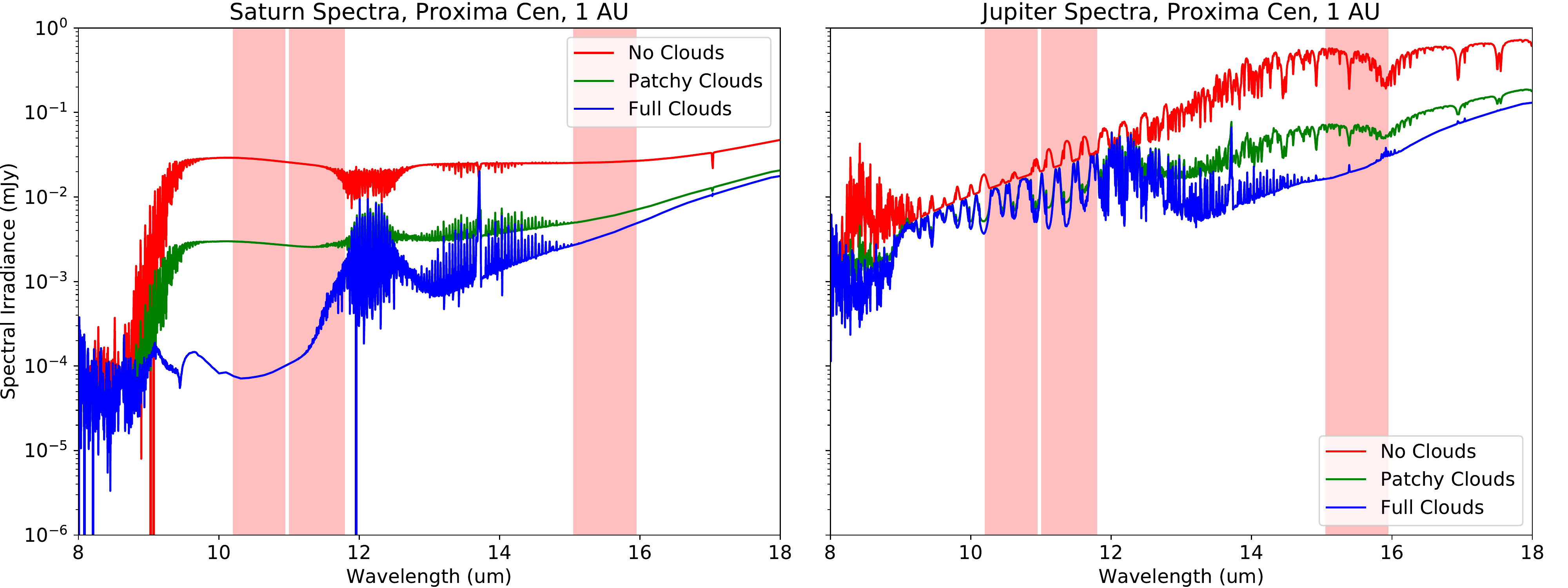}
\caption{Example emission spectra for both Saturn (left) and Jupiter (right), with MIRI coronagraphic filter bandpasses overlaid. Each planet has three spectra --- A default spectrum: with full attenuation due to clouds, a clear spectrum: with no attenuation due to clouds, and a combined spectrum: to simulate patchy clouds partially attenuating planetary emission.}
\label{figure:example_spectra}
\end{figure*}

We used NASA Goddard's Planetary Spectrum Generator \citep[PSG,][]{villanueva2018}, an online radiative-transfer tool, to synthesize planetary spectra given our various stellar, planetary, and observational parameters.
We modeled our simulated planets as mature Jovian-type planets much like Jupiter and Saturn in our own solar system. PSG's built-in atmospheric templates are available for many solar-system objects such as Jupiter and Saturn, as well as a number of generic giant gaseous exoplanet models. Observations of Jupiter and Saturn show variations in cloud cover and atmospheric scattering that have significant effects on planetary infrared emission. Assuming that clouds uniformly block emission from lower layers in the planetary atmosphere, then the projected area of the planet's disk shows differences in brightness depending on whether or not cloud bands are present. Figure \ref{fig:jupiter_gemini} shows the effect of varying cloud bands on Jupiter's brightness in 4.8$\mu$m as seen from Gemini North using the Near-Infrared Imager instrument. Since PSG's atmospheric templates produce 1-dimensional planetary spectra, clouds are represented as a single layer across the entirety of the spectrum, attenuating it as appropriate for the particular molecular species present in the cloud. To emulate the variations in cloud cover that would partially lower the aggregate brightness of the planet's disk, we produced three model spectra: (1) with attenuation due to clouds and hazes, (2) no clouds or hazes, and (3) a combined model simulating partial cloud cover. 
We assumed a somewhat conservative estimate that a Jovian-type planet is likely to have $\sim90\%$ cloud cover over its disk any any point in time. As such, our combined spectrum is a linear sum made up of 10\% of the cloud-free model and 90\% of the cloudy model. This produced a combined spectrum representing a planet with an intermediate brightness to the full- and no-cloud planets. Figure \ref{figure:example_spectra} shows the synthesized planetary spectra used in this work, as well as overlaid MIRI coronagraphic bandpasses. Given the planetary spectra of Saturn and Jupiter, MIRI's bandpasses from 10 to 16 $\mu$m give us the best chance at detecting planets where they are most luminous. 

\subsection{JWST Exposure Simulations}

A suite of software has been developed by the Space Telescope Science Institute for planning observations with JWST. The JWST Exposure Time Calculator is available both as a web interface and a Python library, Pandeia \citep{pontoppidan2016}, enabling simulated observations from any of JWST's observing modes. Pandeia incorporates the most current performance estimates of the JWST instruments, including MIRI as described in the JWST-MIRI special issue of the Publications of the Astronomical Society of the Pacific, of which the relevant papers to this work are \citet{rieke2015} and \citet{boccaletti2015}. In the course of this investigation, we use the Pandeia/ETC MIRI coronagraphic observing mode. For simulations performed here, the Pandeia coronagraphic imaging mode supported a specific subset of coronagraphic operations, outlined in the JWST User Documentation. \citep{jdox2016} The relevant specifications are as follows:

\begin{itemize}
    \item Pandeia supports only PSF reference differential imaging \citep{jdoxetccoro2016}.
    \item PSF reference observations center the reference star in the coronagraphic mask \citep{jdoxetccoro2016}.
    \item For PSF reference subtraction, the user must specify a PSF reference star close in magnitude and stellar type to the science target  \citep{etccorostrat2016}.
    \item Pandeia assumes no difference in the thermal/dynamical state of the observatory between PSF reference observations and science observations  \citep{etccorostrat2016}.
\end{itemize}

Without data from the observatory as it performs in flight, the simulations performed here provide upper limits on planet detectability. Once JWST is launched and commissioning/early release science observations have been conducted there will be sufficient information to update the ETC and other JWST proposal tools with detailed capabilities to simulate actual observatory and instrument performance. \footnote{While this work was underway, the PanCAKE package \citep{girard2018} was released. PanCAKE incorporates more realistic PSF computations and subtractions and provides additional observation modes, including small-grid dithers. As this functionality was not available during the planning and initial simulations of this work, we do not investigate the use of these methods here, but plan to do so in future work.}

Our goal is to sample a wide potential parameter space of giant, nearby exoplanets. Our full set of simulations of a planetary system consists of simulations performed at three orbital distances (1, 2, and 5 AU), in three separate filters ($10.65\mu$m, $11.4\mu$m, and $15.5\mu$m), for three different atmospheric configurations (cloud-free, patchy cloud, and full cloud cover), and bracketed across a range of exposure times (898--22,470 s). Although we use solar-system Jupiter and Saturn spectra for our planetary atmosphere models, our orbital modeling is idealized to represent the best-case configurations for direct imaging. Our planets were simulated to be on circular orbits, with no inclination to the line of sight, and always situated at maximum projected separation from the host star (phase $90^{\circ}$). This allows us to directly correlate the orbital distance of our synthetic targets with projected angular separation. As coronagraphic performance strongly depends on the angular separation of targets and their host stars, tying orbital distance and angular separation is a useful strategy for investigating favorable planetary targets.

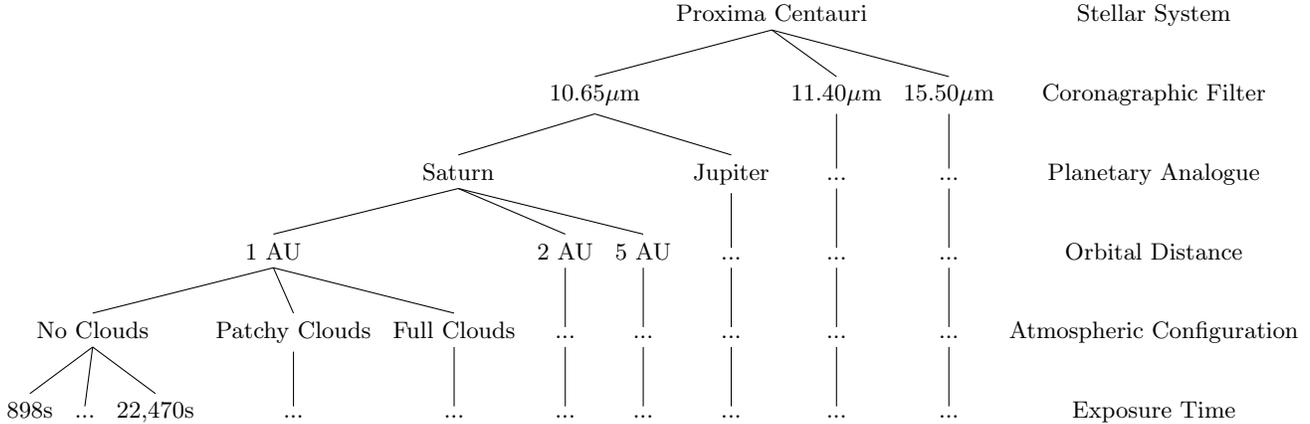
\begin{figure*}[ht]
\begin{tikzpicture}
\begin{scope}[xshift=2in]
\tikzset{edge from parent/.append style={draw=none}, every tree node/.append style={}}
\Tree [. {Stellar System} [. {Coronagraphic Filter} [. {Planetary Analogue} [. {Orbital Distance} [. {Atmospheric Configuration} {Exposure Time} ]]]]]
\end{scope}
\Tree [. {Proxima Centauri} [.10.65$\mu$m 
                                [.Saturn 
                                    [.{1 AU} 
                                        [.{No Clouds} 
                                            898s ... 22,470s ]
                                        [.{Patchy Clouds} ... ]
                                        [.{Full Clouds} ... ]] 
                                    [.{2 AU} [. ... [. ... ]] ] 
                                    [.{5 AU} [. ... [. ... ]] ]] 
                                [.Jupiter [. ... [. ... [. ... ]]] ]]
                            [.11.40$\mu$m [. ... [. ... [. ... [. ... ]]]] ] 
                            [.15.50$\mu$m [. ... [. ... [. ... [. ... ]]]] ]]
\end{tikzpicture}
\caption{Hierarchical structure of possible Pandeia inputs: a depth-first traversal of this tree can be used to populate a JSON file that Pandeia ingests to run a simulated observation. We created a total of over 17,000 individual inputs to span our parameter space.}
\label{figure:etc_inputs}
\end{figure*}

To perform simulations on a full set of planets, expanding along each desired parameter and investigating such planets around a large sample of stellar systems, we elected to use Pandeia. This allowed us to programmatically construct large sets of inputs and parallelize the execution of those simulations. In total, we performed over 17,000 individual simulations (see Fig. \ref{figure:etc_inputs}), so having access to such parallelized simulations was critical.

The basic Pandeia input is a JavaScript Object Notation (JSON) file which defines a hierarchical data structure containing information describing the state of the telescope, the astronomical scene being observed, the observing strategy, the astronomical background, and other relevant observational data. This hierarchical structure simplifies the process of accessing and modifying data in a particular input file, by allowing each field to be directly referenced and changed. If each stellar system is treated as a tree-based structure, a depth-first traversal of nodes in the tree produces a single sequence of parameters, and a path down to each leaf represents a unique, fully-populated input file. Therefore, looping through a list of every star system to be observed can produce a full set of ETC inputs. A section of such an input tree, describing a single star's worth of parameters, is shown in Figure~\ref{figure:etc_inputs}. 

Pandeia requires that astronomical sources be described by either built-in spectral models or user-defined spectra. Our stellar sources can be appropriately described using the available PHOENIX stellar models, which Pandeia uses as inputs to the WebbPSF tool to create a stellar PSF library, and for the planetary sources we used the previously created PSG spectra. These spectra can be directly placed into the corresponding Pandeia input file for use in simulating a JWST observation. Once the input file is complete, Pandeia's {\texttt{perform\_calculation}} function takes the input file, runs the observation simulation, and returns a new JSON object containing the results of the calculation, including total exposure time and SNR.

GSFC's Goddard Private Cloud (GPC) computing resource was used to perform the large set of simulations. Because each simulation is independent of the others, no complicated communication or synchronization is required, and the simplest solution was to run a number of multithreaded virtual machines concurrently to perform as many simulations as possible. All the inputs were created, populated, and stored into a single backing database. Each virtual machine would check the database for an unstarted simulation, pull out that specific input structure, and then start a simulation using the selected input. When a simulation finished execution, the machine would insert the calculated signal-to-noise ratio and the exposure time back into the database for that specific simulation. Each simulation could then be referenced by a unique identifier to immediately find the results. 

\section{Results}
\label{sec:results}
\begin{figure*}[ht!]
\plotone{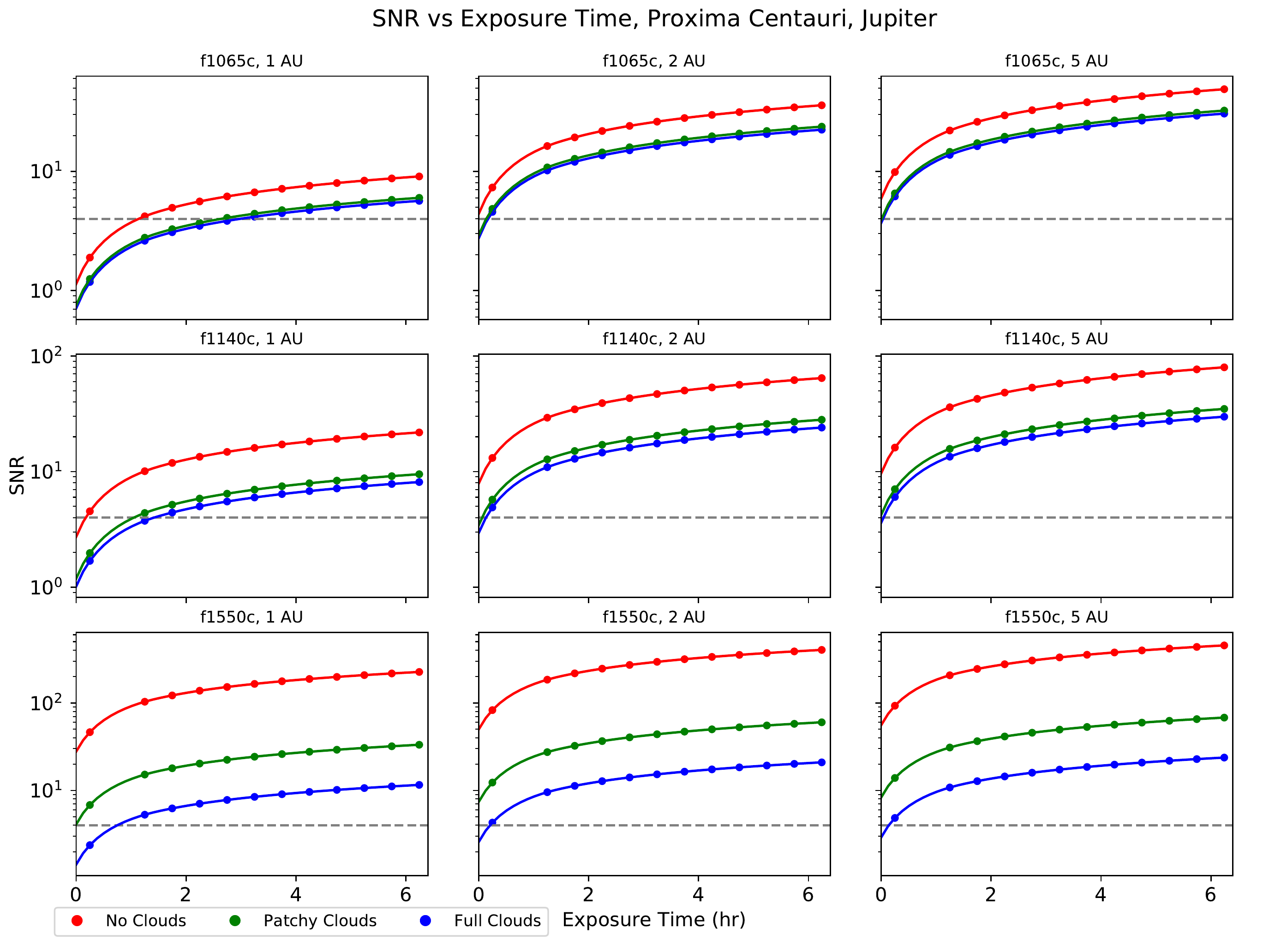}
\caption{SNR/Exposure time plots for Jupiter around Proxima Centauri. Each plot shows three curves: one for each atmospheric simulation (cloud-free, patchy clouds, and full cloud-cover). Plots are shown for each combination of coronagraphic filter and orbital distance. The dashed line shows a detection limit of SNR=4.}
\label{figure:snr_exp_prox_cen_jup}
\end{figure*}

\begin{figure*}[ht!]
\plotone{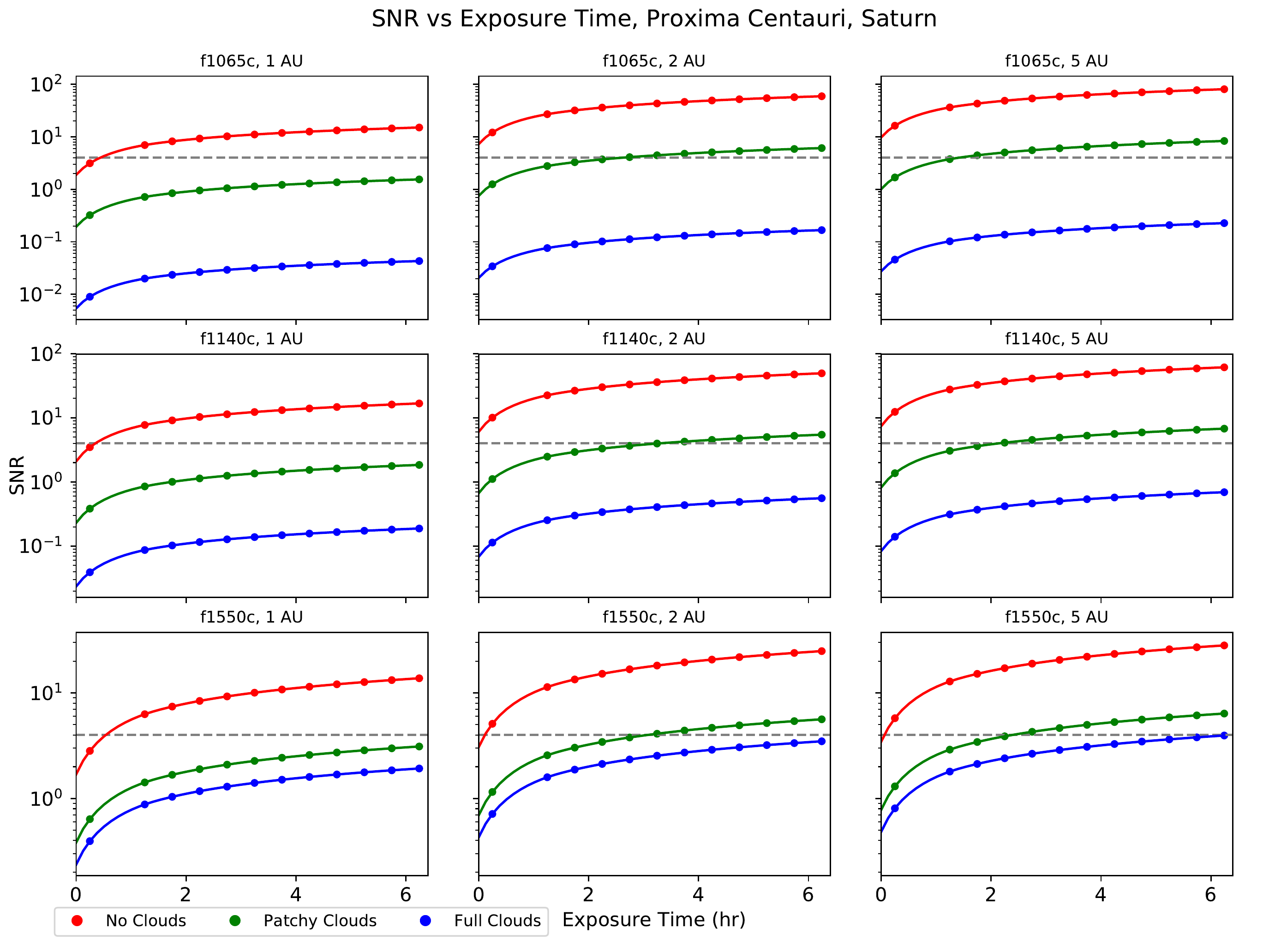}
\caption{Same as Figure \ref{figure:snr_exp_prox_cen_jup}, but for Saturn. Saturns are much more challenging to detect than Jupiters -- only the cloud-free model can be seen in all simulated observations. Partial cloudy planets can be detected at 2 and 5 AU in all three filters, but the cloudy model is only detectable in the most optimistic scenarios.}
\label{figure:snr_exp_prox_cen_sat}
\end{figure*}

We ran our simulations for each star/planet/orbit/filter combination, and created an individual SNR/Exposure time plot. We investigated exposure time by trying to minimize saturation in each up-the-ramp exposure, and then bracketing between a number of exposures. This yielded a minimum integration time of 898.8 seconds and a maximum of 6.24 hours. For each combination we plotted three SNR/Exposure time curves: one for each atmospheric configuration. The SNR/Exposure time curves were found by interpolating between the discrete points using a quadratic approximation of the SNR behavior. We classed a planet as detectable if the SNR was above 4 at the maximum integration duration of 6.24 hours. Figures \ref{figure:snr_exp_prox_cen_jup} and \ref{figure:snr_exp_prox_cen_sat} show an example selection of results for one star in our sample, Proxima Centauri.

Proxima Centauri is the closest star to Earth, and also is on the lower end of the temperature range of our sample. As such, sample planets around Proxima Centauri are the highest SNR detections we have. 
Fig. \ref{figure:snr_exp_prox_cen_jup} shows Jupiters around Proxima Centauri, and we find that only the 1 AU planets take any significant time to be able to detect. Our worst-case observation is Jupiter at 1 AU in the $10.65\mu$m filter, and it is detectable for all cloud configurations within about 3.5 hours. Fig. \ref{figure:snr_exp_prox_cen_sat} shows that for Saturns, things are somewhat more challenging, with only the brightest planets detectable within our maximum exposure time of 6.24 hours. No Saturn is actually detectable at all atmospheric configurations at our maximum exposure time; the 6.24 hour, $15.5\mu$m, cloudy Saturn misses a 4 SNR detection by 0.05.

\begin{figure*}[ht!]
\plotone{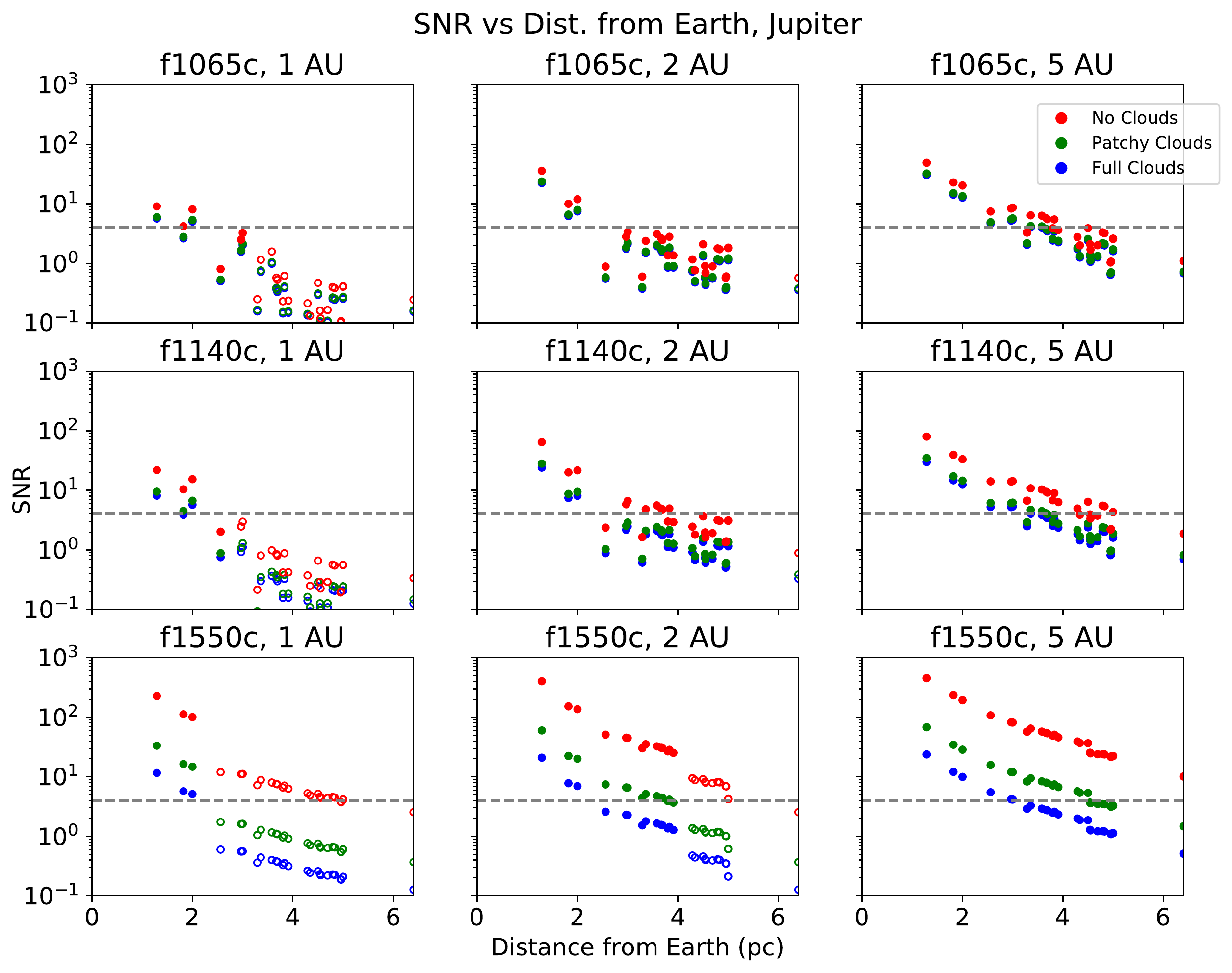}
\caption{Maximum SNR/Distance plots for Jupiter-mass planets. Each point shows the maximum SNR observation for the system at that distance. Open circles represent systems within the IWA for that specific coronagraph. Again, the dashed line shows a detection limit of SNR=4.}
\label{figure:jupiter_max}
\end{figure*}

\begin{figure*}[ht!]
\plotone{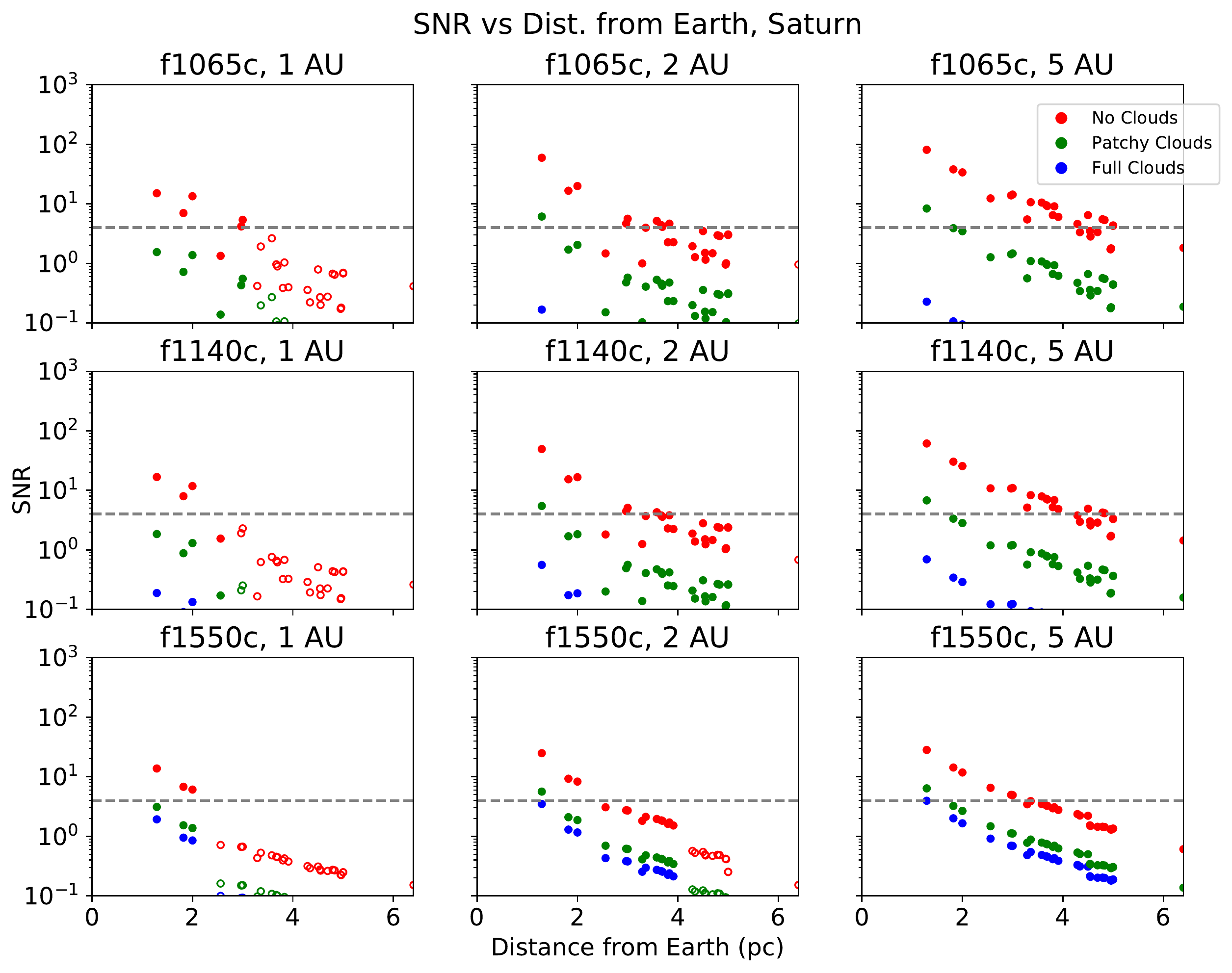}
\caption{Same as Figure \ref{figure:jupiter_max}, but for Saturn.}
\label{figure:saturn_max}
\end{figure*}

Figures \ref{figure:jupiter_max} and \ref{figure:saturn_max} show how planet detectability changes with system distance. Each point plotted in these two figures represents the max-SNR/longest duration exposure (6.24hr) observation for that specific planet. Open circles represent planets within the inner working angle for that particular coronagraph. Ordinarily, it would be difficult to make any sort of determination about potential detections within the IWA. However, some of our sub-IWA simulated planets produce detections near SNR=10, particularly for Jupiters in 15.50$\mu$m at 1 and 2 AU. In Figures \ref{figure:jupiter_max} and \ref{figure:saturn_max}, drops in SNR due to the IWA can be clearly seen. 

Each star in our sample has at least one detectable planet configuration, for a total of 300 detectable planets. 222 of these planets are Jupiters, and 78 are Saturns. 77 planets were detectable in the $10.65\mu$m filter, 88 in $11.4\mu$m, and 135 in $15.5\mu$m. 199 planets are cloudless, 66 have patchy clouds, and 35 are fully clouded. 56 planets are detectable at 1 AU, 87 are detectable at 2 AU, and 157 are detectable at 5 AU. For the most favorable system configuration (cloud-free Jupiters at 5 AU in $15.5\mu$m), we can detect at least one planet within our maximum 6.24 hour exposure time around every star in our sample. As shown in Figures \ref{figure:jupiter_max} and \ref{figure:saturn_max}, system distance is the most fundamental factor affecting planet detectability. We also find that some simulated planets may be detectable inside the IWA’s of the MIRI coronagraphs. More details are provided in Appendix \ref{appendix:sub_iwa}.

As shown in Figure \ref{figure:jupiter_max}, MIRI can be used to detect completely cloud-covered Jupiters around stars within 3 pc, partially cloudy planets around stars within 5 pc, and cloud-free well beyond 6.4 pc. Saturns are more challenging, as we are unable to detect any fully-clouded Saturns, and can only detect partially clouded Saturns around Proxima Centauri. However, cloud-free Saturns can be detected around stars out to 4 and 5 parsecs. Because our simulations only investigated planets within 5 AU of their host stars, these results present upper bounds on the exposure times needed to detect planets further than 5 AU. Sections 4 and 5 of \citet{boccaletti2015} show the contrast curves and simulated system performance (Figures 9 and 10), as well as some simulated MIRI coronagraphic observations of Jovian-type exoplanets and the HR8799 system. As the simulated exoplanetary targets (especially in the HR 8799 case) move further away from the center of the coronagraphic image, they become more detectable as they are less attenuated by the coronagraph. Following this, if we observe a system and target a particular planet, say a Jupiter at 1 AU, we are necessarily sensitive to potentially undiscovered outer companions, provided their projected separation from the star is large enough. This full parameter space where JWST MIRI is sensitive to more massive and more widely separated planets is shown as the dashed box in Figure~\ref{figure:planets}. More detailed maximum SNR/exposure time information is available in Appendix \ref{table:max_snrs}, and the full maximum SNR/exposure time table is available in machine-readable format.

\section{Discussion}

\subsection{Known Potential Targets - GJ 832 b}
\label{sec:known_targets}
One of the stars in our sample population, GJ 832, is known to have a Jovian exoplanet \citep{bailey2009} with a minimum mass of $\sim$0.68 $M_{J}$ at an orbital distance of $\sim$3.56 AU \citep{wittenmyer2014}. GJ~832 b was discovered through RV measurements taken over a ten-year period and is not known to transit its host star. Given its minumum mass and wide maximum angular separation ($\sim$0.69$''$), it should be a tempting target for direct detection efforts like ours. The inclination of the system is unknown, but by assuming $45^{\circ}$ inclination, and estimating the system properties (i.e. nearly circular, with projected orbital radius $\sim$0.69$''$) we derive a minimum separation close to the projected boundaries of the IWA of the 15$\mu$m coronagraph. Detecting GJ~832~b with this instrument would therefore be challenging or impossible for about half of GJ~832~b's $\sim$3600 day orbit. However, given the planet only has a mass lower bound of M sin(i), if its true orbital inclination is close to $45^{\circ}$, its true mass could be closer to $\sim1 M_{J}$. To some extent, this would make GJ~832~b a slightly better imaging target for the portions of the orbit where it is observable.

\begin{figure*}[ht!]
\plotone{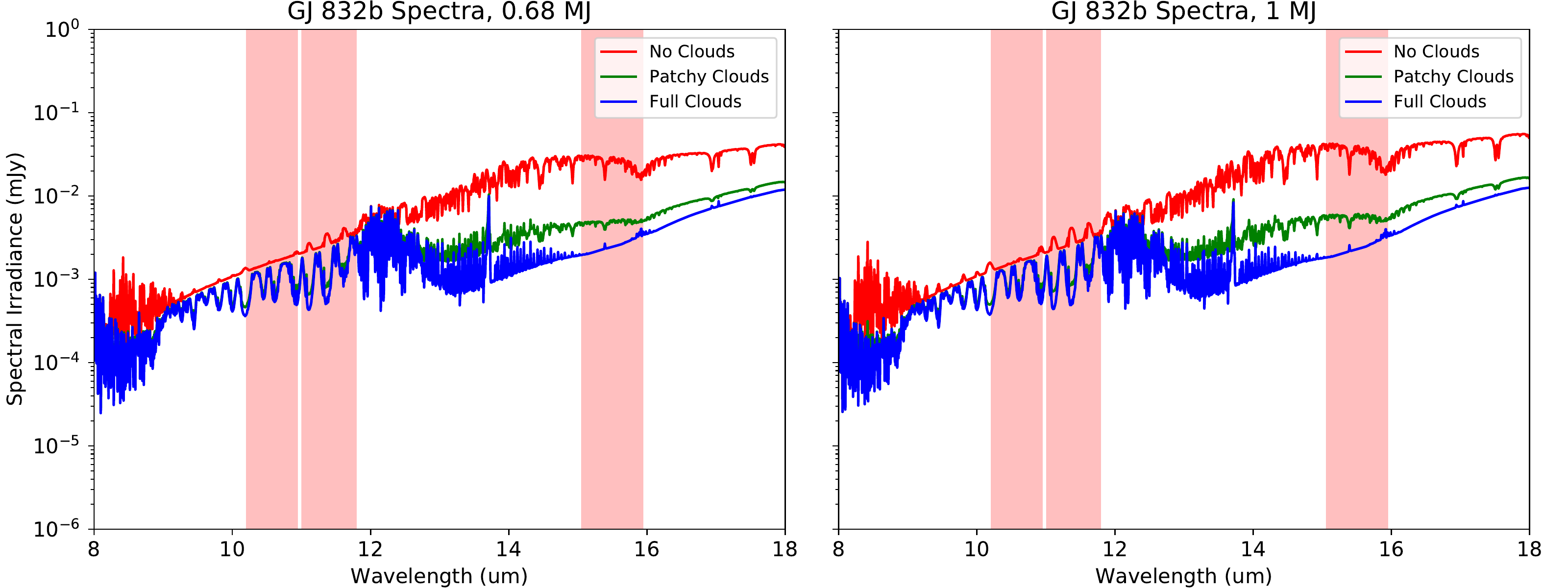}
\caption{GJ 832b sample spectra, using Jupiter-type atmospheres at $0.68 M_{J}$ and $1 M_{J}$}
\label{fig:gj832b_specs}
\end{figure*}

We simulated the detectability of this system, following the process used for the rest of our simulations. We produced a set of PSG spectra (shown in Figure~\ref{fig:gj832b_specs}) and Pandeia inputs using the known parameters of the planet, and substituted modeled planetary parameters for GJ~832~b where needed. Since we only have an estimated mass for GJ~832~b, we used the \textsc{forecaster} Python package \citep{forecaster} to estimate the radius of the planet. For each mass (0.68 $M_{J}$ and 1.0 $M_{J}$) the mean predicted radii were both $\sim1.25 R_{J}$. Using Jupiter-type atmospheres again, we created spectra for both masses, placed those into the relevant Pandeia inputs, and simulated JWST observations using MIRI in the $15.5 \mu$m filter.

\begin{figure*}[ht!]
\plotone{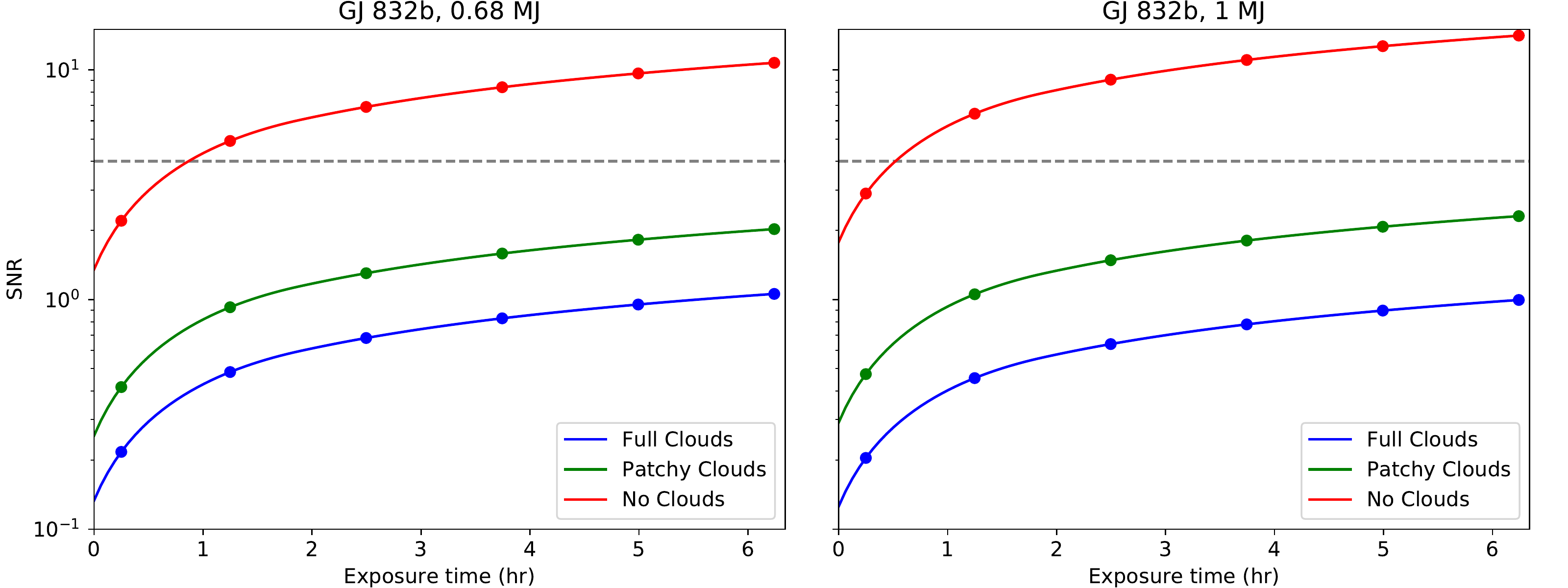}
\caption{GJ 832b SNR vs Exposure time, seen in 15.5$\mu$m}
\label{fig:gj832b_snr}
\end{figure*}

As in previous simulations, we found that our potential detections are split between the brighter and dimmer atmospheric configurations. We found that for both the $0.68 M_{J}$ and $1 M_{J}$ cases, we were only able to detect the planets if they did not have significant cloud cover. Due to the significant gaps in our understanding of the system, we feel comfortable in designating GJ~832~b a potentially detectable planet in the $15.5\mu$m filter. Figure~\ref{fig:gj832b_snr} shows SNR vs. Exposure time curves for our GJ 832b simulations. 

As of now, GJ~832~b is the only known potential target in our sample of nearby M-stars. However, given the estimates on the occurrence rates of Jovian-type planets around M-dwarfs, there may be another undiscovered planet or two within our stellar sample. Even if this is not the case, potentially imaging GJ~832~b would be an unprecedented feat, which could lead to a new era of detecting and potentially characterizing long-period giant exoplanets.

\subsubsection{MCMC orbital modeling}
\begin{figure*}[ht!]
\plotone{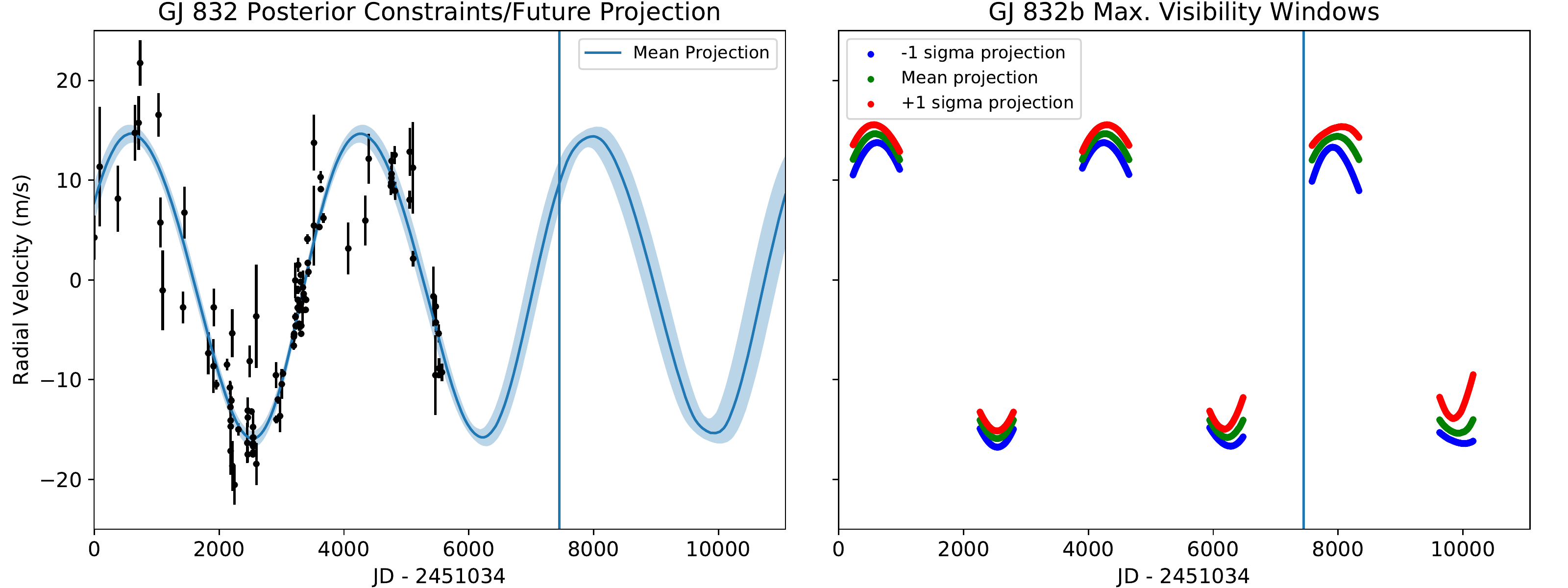}
\caption{Historical fit and forward projections of GJ~832~b radial velocity and maximum visibility windows, with $\pm 1 \sigma$ estimates. The orbital parameters were modeled using the {\textsc{exoplanet}} MCMC package, and the vertical lines indicates the date 01/01/2019.}
\label{fig:gj832b_post}
\end{figure*}

Since GJ~832~b is only known from radial velocity data, and that data is only complete up until 2010, we decided to predict when the planet would be observable in the future by JWST. We used the Monte Carlo simulation tools in the {\textsc{exoplanet}} \citep{exoplanet} Python package to model the orbit of GJ~832~b given some of the previously determined orbital parameters from past work. Projecting the orbit into the future, we found that GJ~832~b would be most observable near September 2025 (JD 2460930). The results of the orbital modeling are shown in Figure~\ref{fig:gj832b_post}, along with the predicted maximum visibility windows. However, this analysis is not quite sufficient, since potential exoplanetary targets are liable to be outside JWST's visibility zones for some portion of the year. Nearby stars may also affect the quality of JWST observations. Therefore, we also evaluated GJ~832~b's visibility and potential contamination using the STScI's Exoplanet Characterization Toolkit's Contamination and Visibility Calculator (ExoCTK). According to the ExoCTK output, the system is observable without background stellar contamination from May to July, between JWST observatory V3 position angles of about $240^{\circ}$ to $290^{\circ}$. There is also a period of visibility from September to November 2025 at lower position angles of $30^{\circ}$ to $70^{\circ}$, but that window has moderate background contamination. Comparing this with STScI's JWST Coronagraph Visibility Tool, we find that GJ~832 would be visible in quadrants 1, 2, and 4 of MIRI's $15.5\mu$m coronagraph across the feasible range of position angles. Ideally we would also observe this system in December 2024 (JD 2460661) and June 2026 (JD 2461199) to further constrain the planet's orbit, with additional observations bracketing the point of maximum projected separation, about a year before and after. Due to the restrictions from the spacecraft's orbit and stellar background, we instead suggest that the three characterizing observations be made near June 2024, 2025, and 2026 (JD 2460469, 2460834, and 2461199).

\subsection{Synergy with Gaia}
Combining Gaia astrometry, radial velocity, and direct imaging observations allows for refining of currently-theorized orbits and some characterization of system architectures \citep{brandt2019}. Currently-known radial velocity planets can be further characterized with direct imaging, but with only a few imaging epochs may still be somewhat unconstrained in inclination and mass. Long-baseline astrometric observations would be critical to determining inclination, which would then refine mass estimates from radial velocity, which could then break the degeneracy between mass and orbital phase in direct imaging observations.

Again, GJ~832~b  functions as a useful test system. The planet is known from radial velocity observations, and the system will be observed by Gaia 75 times over the entirety of its main mission. If the planet is recovered only by JWST coronagraphic imaging, it will likely be difficult to further characterize the system with only a few observations. If analysis of the Gaia data is able to characterize the planet's orbit and determine its inclination, then the planet's mass will be apparent and the planet can be characterized more fully by the direct imaging data. Any other potentially detected exoplanets in our sample may also be characterized in a similar way.

\subsection{Example Observing Programs}

\begin{deluxetable*}{lcclccccr}
\caption{Optimized Observing Targets}
\label{table:opt_obs}
\tablehead{\colhead{Star} & \colhead{Filter} & \colhead{Planet} & \colhead{Atmosphere} & \colhead{SMA} & \colhead{Separation} & \colhead{Separation} & \colhead{SNR} & \colhead{Exposure} \\ 
\colhead{} & \colhead{} & \colhead{} & \colhead{} & \colhead{(AU)} & \colhead{(arcsec)} & \colhead{($\lambda / D$)} & \colhead{} & \colhead{(seconds)} } 
\startdata
Proxima Cen & f1065c & Jupiter & Full Clouds & 2 & 1.55 & 4.70 & 4.57 & 898.8 \\
Wolf 359 & f1550c & Jupiter & Patchy Clouds & 2 & 1.00 & 2.04 & 4.10 & 898.8 \\
GJ 699 & f1550c & Jupiter & Patchy Clouds & 2 & 1.10 & 2.24 & 4.59 & 898.8 \\
GJ 411 & f1550c & Jupiter & Patchy Clouds & 5 & 1.95 & 3.99 & 7.20 & 4494.0 \\
Ross 154 & f1550c & Jupiter & Patchy Clouds & 5 & 1.68 & 3.44 & 5.46 & 4494.0 \\
Ross 248 & f1550c & Jupiter & Patchy Clouds & 5 & 1.67 & 3.40 & 5.42 & 4494.0 \\
GJ 887 & f1550c & Jupiter & Patchy Clouds & 5 & 1.52 & 3.10 & 4.51 & 6291.6 \\
Ross 128 & f1550c & Jupiter & Patchy Clouds & 5 & 1.49 & 3.04 & 4.29 & 4494.0 \\
DX Cnc & f1550c & Jupiter & Patchy Clouds & 5 & 1.40 & 2.85 & 4.51 & 6291.6 \\
GJ 1061 & f1550c & Jupiter & Patchy Clouds & 5 & 1.36 & 2.78 & 4.28 & 6291.6 \\
YZ Ceti & f1550c & Jupiter & Patchy Clouds & 5 & 1.36 & 2.77 & 4.22 & 6291.6 \\
Luyten's & f1550c & Jupiter & Patchy Clouds & 5 & 1.32 & 2.69 & 4.35 & 8089.2 \\
GAT 1370 & f1550c & Jupiter & Patchy Clouds & 5 & 1.31 & 2.66 & 4.52 & 8089.2 \\
Kapteyn's & f1550c & Jupiter & Patchy Clouds & 5 & 1.28 & 2.61 & 4.08 & 8089.2 \\
GJ 628 & f1550c & Jupiter & Patchy Clouds & 5 & 1.17 & 2.38 & 4.17 & 11684.4 \\
GJ 1 & f1550c & Jupiter & Patchy Clouds & 5 & 1.15 & 2.35 & 4.21 & 13482.0 \\
LHS 292 & f1550c & Jupiter & Patchy Clouds & 5 & 1.11 & 2.27 & 4.18 & 13482.0 \\
\enddata
\end{deluxetable*}

We can use our SNR/Exposure time data to calculate the length of some sample observing programs. Because each system we simulated has at least one detectable planet (of any atmospheric configuration), if we assume each planet takes the maximum exposure time to detect, a full survey of our stellar sample could recover every detectable planet within $6.24 \times 27 = 168.48$ hr of integration time. We can refine this selection process to identify a more modest program. Because we target giant gaseous exoplanets, we expect that our cloud-free atmospheric models are overly optimistic for any of the actual exoplanets likely to be observed in this program. We select detectable planets with patchy- and fully-clouded atmospheres and select the single minimum-time detection for each star. Of our original 27-star sample, 10 stars only had detectable planets with the clear atmosphere model, so we exclude these systems from this program. Among the remaining 17 systems, the detectable planets requiring the least integration time were all Jupiters. Except for the Proxima Centauri system, all the planets had patchy clouds, and were best-observable in $15.5\mu$m. A Jupiter with full cloud cover orbiting Proxima Centauri was best-observable in $10.65\mu$m at 2 AU. Most of the planets in this ersatz program were at 5 AU. Adding up the individual exposure times needed to detect these planets yields a total program integration time of $\sim30.2$ hours. The specific targets of such a program are listed in Table \ref{table:opt_obs}. This optimized program closely follows the general trends of the results of the full dataset, that is, the most detectable planets are Jupiter-type planets orbiting at 5 AU, observed in the $15.5\mu$m coronagraphic filter. 

Depending on how efficiently the telescope overhead can be optimized, this program may fit into the time limits of a JWST medium-size GO proposal, originally projected to be $25 - 75$hr in total time allocation. Three observations in our sequence are $>10,000$ seconds, so removing those observations would shrink the required integration time to $\sim19.5$ hours, which would almost certainly fit into a JWST medium-size GO proposal, including telescope overhead. Notably, the subset of stars that would be observed in this program does not include GJ 832, since all detectable planets around that star were cloud-free, both in our full set of simulations and in our modeling of the known planet GJ~832~b. More time could be devoted to searching for that specific planet, but that would likely necessitate a longer observing program. The observations shown here are near the minimum-possible time required to detect planets around the listed stars, and the planets being detected are the most optimal simulated configurations for detection. No such planets are known to fit these parameters, so a realistic survey of these stars for detectable planets should likely observe each system for the maximum exposure time. This would bring the total integration time of this 17-star survey to 106.08 hours, solidly in the large-size GO Program category.

\section{Conclusions}
We present an investigation into the feasibility of JWST's Mid-Infrared Instrument for the direct imaging of cold Jovian-type planets around nearby M-dwarfs. Any planets directly imaged in this sample would be the nearest to Earth, as well as the lowest-mass and closest-orbiting of any giant planets yet directly imaged. MIRI's sensitivity to these $0.3-1$M$_{Jup}$ planets at projected separations of 1-5 AU also implies that it would be sensitive to similar (or more massive) planets at wider separations where there are more favorable contrast ratios (see Fig.~\ref{figure:planets}). In addition, more detailed simulations of the known planet GJ~832~b reveal that it could potentially be recovered with MIRI imaging. GJ~832~b would be the second-closest and likely the least massive exoplanet yet to be imaged.

In order to maximize the chances of such detections, we suggest that future observing programs utilize MIRI's 15.5$\mu$m coronagraphic filter. Although the 15.5$\mu$m coronagraph has the largest inner working angle of MIRI's 4QPM coronagraphs, the slightly reduced sample size is worth the increase in SNR associated with searching for planets with peak emission nearer 15$\mu$m. However, if a planet does lie within the coronagraph's IWA, it may still be detectable provided it is sufficiently bright (see~\ref{appendix:sub_iwa}). This allows for some flexibility in the sample of stellar systems used as potential targets, especially if any new long-period Jovian-mass planets are discovered before JWST's launch or early in the mission. Given that in most cases we can detect our desired planets within 6.24 hours of observing time (excluding telescope operations overhead), a full survey of all 27 stars in our sample would be possible in under 170 hours of observing time for one filter.

As this work has demonstrated the feasibility of using the MIRI coronagraphic imaging modes to detect cold Jovian planets around some of the nearest M-dwarfs, it naturally motivates future work on a full survey simulation. Such an expanded simulation would take into account an input planet population rooted in measured occurrence rates, more varied planet parameters, more accurate stellar spectra and PSF modeling, randomized orbits, and detailed estimates of telescope and coronagraphic inefficiencies. The goal would be to obtain realistic completeness estimates and investigate the full sensitivity parameter space for a larger stellar host sample.

JWST coronagraphic imaging will also be supplemented in the years after launch by the Wide-Field Infrared Survey Telescope (WFIRST). WFIRST's microlensing survey will be extremely sensitive to Jovian-mass and smaller planets at intermediate orbital distances of 1-10 AU (see Fig. 9 in \citet{penny2019}), and WFIRST will also be capable of high-contrast coronagraphic imaging from 575 -- 825 nm at IWAs down to $0.15''$ \citep{mennesson2018}. Jovian planet statistics from the microlensing sample could inform future JWST imaging campaigns, and WFIRST direct imaging will usher in a new era of exoplanet characterization through reflected light spectroscopy.

Further in the future, the Astro2020 Decadal Survey large mission concept proposals contain a number of observatory concepts capable of high-contrast coronagraphy and direct imaging. The Large UV/Optical/IR Surveyor (LUVOIR) and the Habitable Exoplanet Observatory (HabEx) are intended to directly image differing populations of exoplanets in varied UV, optical, and infrared bands. LUVOIR's Extreme Coronagraph for Living Planetary Systems (ECLIPS) will be capable of coronagraphic imaging in the near-UV, optical, and near-IR \citep[0.2--2 $\mu$m,][]{luvoir2019}. HabEx will be capable of coronagraphy in the optical and near-IR \citep[0.45--1.8 $\mu$m,][]{habex2019}. These missions are intended to discover and characterize exo-Earth type planets, and represent the next great frontier in space-based direct imaging.

\vspace{5mm}

\acknowledgments
The authors would like to thank Michael Moore and the rest of the Goddard Private Cloud team for their support and assistance. This research has made use of NASA’s Astrophysics Data System Bibliographic Services; the NASA Exoplanet Archive, which is operated by the California Institute of Technology, under contract with the National Aeronautics and Space Administration under the Exoplanet Exploration Program; the SIMBAD database, operated at CDS, Strasbourg, France; and the RECONS 10 Parsec Sample, compiled by Georgia State University, Atlanta, Georgia, USA (http://www.astro.gsu.edu/RECONS/index.htm). This work was supported by the Sellers Exoplanet Environments Collaboration (SEEC) at NASA's Goddard Space Flight Center. We would like to thank Michael McElwain for enlightening discussions that greatly improved this manuscript. We also thank the anonymous reviewer for their helpful feedback.

This research was conducted on the historical territory of the Piscataway people.

\software{
 Matplotlib \citep{matplotlib}, 
 NumPy \citep{numpy},
 Pandeia \citep{pontoppidan2016}, 
 forecaster \citep{forecaster}, 
 exoplanet \citep{exoplanet},
 Astropy \citep{astropy13,astropy18},
 Theano \citep{theano}
 PyMC3 \citep{PyMC3}
}
\facility{Exoplanet Archive}

\pagebreak

\appendix
\section{Prospects for Sub-Inner Working Angle Planet Detections}
\label{appendix:sub_iwa}
While the IWA is defined as the 50\% transmission radius of the MIRI coronagraphs, the shape of the radial transmission curves inside the IWA determines whether sub-IWA imaging is feasible at all. Figure 6 in \citet{boccaletti2015} shows the radial transmission curves for both the 4QPM and Lyot MIRI coronagraphs. For the Lyot coronagraph, the radial transmission curve falls extremely steeply from near 90\% transmission at $\sim 4 \lambda / D$, to 50\% transmission at $\sim 3 \lambda / D$, 10\% at $\sim 2.75 \lambda / D$, and 5\% at $\sim 2 \lambda / D$. As the IWA of the Lyot coronagraph is 2.16$''$, this makes recovery of targets effectively impossible within $\sim2''$ of the center of the image. For the 4QPM coronagraphs, the radial transmission curve is not as steep as the Lyot coronagraphs. 100\% relative transmission occurs at $\sim 3 \lambda / D$, 50\% at $\sim 1 \lambda/D$, and 20\% transmission at $\sim 0.5 \lambda/D$. Given the sub-IWA transmission curves, we adopt the $0.5 \lambda / D$, 20\% transmission limit as an inner bound for where sub-IWA detections are possible. For each of the $10.65\mu$m, $11.4\mu$m, and $15.5\mu$m coronagraphs, this $0.5 \lambda / D$ limit corresponds to $0.165''$, $0.18''$, and $0.245''$, respectively. Given that we have simulated cold gas-giants at relatively small projected separations (1, 2, and 5 AU) around the nearest stars, some are likely to fall within the IWAs of the coronagraphs.  We expect that planets in this regime will be detectable provided they are bright enough to overcome the sub-IWA attenuation by the coronagraphs.

In addition, because the PSF of the planetary companion is likely to be entirely encompassed by the stellar host's PSF, signal from the planet can only be recovered when the PSF subtraction does not entirely zero out the pixels containing the image of the planet. As such, in the course of such sub-IWA observations, care must be taken to prevent the camera from being saturated in both the target image and the PSF reference image. Otherwise, the standard coronagraphic imaging process and reference PSF subtraction can be used to recover sub-IWA planets, provided they are bright enough to overcome the steep attenuation of the coronagraph inside this region. 

In our suite of simulations, 33 unique planet configurations are detectable inside the IWA of the MIRI coronagraphs. These are all Jupiter-type planets, with no clouds, and were only observable with the $15.5\mu$m coronagraph. Since our planetary atmospheres are brightest at $15.5\mu$m without clouds, we do expect that these planets would be most detectable at this wavelength. Also, as the $15.5\mu$m coronagraph has the widest IWA, so we do expect that there would be a certain range of stellar host distances where planets with smaller orbital distances would lie within the IWA and the wider orbital distances may be outside the IWA. In these cases, planets are detectable around 23 of our stellar hosts, ranging from 2.97 - 6.4 pc away.

Further coronagraphic imaging techniques, like small-grid dithering \citep{lajoie2016}, are available to optimize the PSF subtraction. In situations where the PSF reference target acquisition is uncertain, the small-grid dither technique can significantly improve PSF reference subtraction performance. This adds some time expense to JWST coronagraphic operations, as the reference star needs to be more thoroughly observed. JWST commissioning and preliminary science observations may show that these techniques will be necessary for imaging challenging coronagraphic targets, and sub-IWA targets in particular. Although JWST in-flight pointing and thermal performance is not yet known, current performance requirements indicate that some of the more marginally detected targets, especially the aforementioned 33 sub-IWA planets, may be rendered undetectable. Future work will be needed to better quantify this and should incorporate the updated PanCAKE package and less optimistic image post-processing, such as the inclusion of pointing drift between science targets and PSF references.

\section{Exposure times and Signal-to-Noise ratio for simulated planets orbiting nearby cool stars}
\label{appendix:exp_time}

\startlongtable
 \begin{deluxetable*}{lcclccrr}
 \tablecaption{Planet SNR at maximum exposure (6.24 hr), for planets placed around Proxima Centauri}
 \tablehead{\colhead{Star} & \colhead{Filter} & \colhead{Planet} & \colhead{Atmosphere} & \colhead{SMA} & \colhead{Separation} & \colhead{Separation} & \colhead{SNR} \\ 
\colhead{} & \colhead{} & \colhead{} & \colhead{} & \colhead{(AU)} & \colhead{(arcsec)} & \colhead{($\lambda / D$)} & \colhead{} } 
\startdata
Proxima Cen & f1065c & Jupiter & Full Clouds & 1 & 0.78 & 2.35 & 5.65 \\
Proxima Cen & f1065c & Jupiter & Full Clouds & 2 & 1.55 & 4.70 & 22.36 \\
Proxima Cen & f1065c & Jupiter & Full Clouds & 5 & 3.88 & 11.75 & 30.50 \\
Proxima Cen & f1065c & Jupiter & Patchy Clouds & 1 & 0.78 & 2.35 & 6.00 \\
Proxima Cen & f1065c & Jupiter & Patchy Clouds & 2 & 1.55 & 4.70 & 23.71 \\
Proxima Cen & f1065c & Jupiter & Patchy Clouds & 5 & 3.88 & 11.75 & 32.34 \\
Proxima Cen & f1065c & Jupiter & No Clouds & 1 & 0.78 & 2.35 & 9.06 \\
Proxima Cen & f1065c & Jupiter & No Clouds & 2 & 1.55 & 4.70 & 35.84 \\
Proxima Cen & f1065c & Jupiter & No Clouds & 5 & 3.88 & 11.75 & 48.86 \\
Proxima Cen & f1065c & Saturn & Full Clouds & 1 & 0.78 & 2.35 & 0.04 \\
Proxima Cen & f1065c & Saturn & Full Clouds & 2 & 1.55 & 4.70 & 0.17 \\
Proxima Cen & f1065c & Saturn & Full Clouds & 5 & 3.88 & 11.75 & 0.23 \\
Proxima Cen & f1065c & Saturn & Patchy Clouds & 1 & 0.78 & 2.35 & 1.55 \\
Proxima Cen & f1065c & Saturn & Patchy Clouds & 2 & 1.55 & 4.70 & 6.12 \\
Proxima Cen & f1065c & Saturn & Patchy Clouds & 5 & 3.88 & 11.75 & 8.35 \\
Proxima Cen & f1065c & Saturn & No Clouds & 1 & 0.78 & 2.35 & 15.07 \\
Proxima Cen & f1065c & Saturn & No Clouds & 2 & 1.55 & 4.70 & 59.40 \\
Proxima Cen & f1065c & Saturn & No Clouds & 5 & 3.88 & 11.75 & 80.81 \\
Proxima Cen & f1140c & Jupiter & Full Clouds & 1 & 0.78 & 2.15 & 8.12 \\
Proxima Cen & f1140c & Jupiter & Full Clouds & 2 & 1.55 & 4.31 & 24.03 \\
Proxima Cen & f1140c & Jupiter & Full Clouds & 5 & 3.88 & 10.77 & 29.82 \\
Proxima Cen & f1140c & Jupiter & Patchy Clouds & 1 & 0.78 & 2.15 & 9.49 \\
Proxima Cen & f1140c & Jupiter & Patchy Clouds & 2 & 1.55 & 4.31 & 28.07 \\
Proxima Cen & f1140c & Jupiter & Patchy Clouds & 5 & 3.88 & 10.77 & 34.83 \\
Proxima Cen & f1140c & Jupiter & No Clouds & 1 & 0.78 & 2.15 & 21.80 \\
Proxima Cen & f1140c & Jupiter & No Clouds & 2 & 1.55 & 4.31 & 64.34 \\
Proxima Cen & f1140c & Jupiter & No Clouds & 5 & 3.88 & 10.77 & 79.73 \\
Proxima Cen & f1140c & Saturn & Full Clouds & 1 & 0.78 & 2.15 & 0.19 \\
Proxima Cen & f1140c & Saturn & Full Clouds & 2 & 1.55 & 4.31 & 0.56 \\
Proxima Cen & f1140c & Saturn & Full Clouds & 5 & 3.88 & 10.77 & 0.69 \\
Proxima Cen & f1140c & Saturn & Patchy Clouds & 1 & 0.78 & 2.15 & 1.84 \\
Proxima Cen & f1140c & Saturn & Patchy Clouds & 2 & 1.55 & 4.31 & 5.45 \\
Proxima Cen & f1140c & Saturn & Patchy Clouds & 5 & 3.88 & 10.77 & 6.76 \\
Proxima Cen & f1140c & Saturn & No Clouds & 1 & 0.78 & 2.15 & 16.71 \\
Proxima Cen & f1140c & Saturn & No Clouds & 2 & 1.55 & 4.31 & 49.27 \\
Proxima Cen & f1140c & Saturn & No Clouds & 5 & 3.88 & 10.77 & 61.08 \\
Proxima Cen & f1550c & Jupiter & Full Clouds & 1 & 0.78 & 1.58 & 11.57 \\
Proxima Cen & f1550c & Jupiter & Full Clouds & 2 & 1.55 & 3.16 & 20.93 \\
Proxima Cen & f1550c & Jupiter & Full Clouds & 5 & 3.88 & 7.91 & 23.74 \\
Proxima Cen & f1550c & Jupiter & Patchy Clouds & 1 & 0.78 & 1.58 & 33.24 \\
Proxima Cen & f1550c & Jupiter & Patchy Clouds & 2 & 1.55 & 3.16 & 60.01 \\
Proxima Cen & f1550c & Jupiter & Patchy Clouds & 5 & 3.88 & 7.91 & 67.90 \\
Proxima Cen & f1550c & Jupiter & No Clouds & 1 & 0.78 & 1.58 & 225.57 \\
Proxima Cen & f1550c & Jupiter & No Clouds & 2 & 1.55 & 3.16 & 402.80 \\
Proxima Cen & f1550c & Jupiter & No Clouds & 5 & 3.88 & 7.91 & 453.29 \\
Proxima Cen & f1550c & Saturn & Full Clouds & 1 & 0.78 & 1.58 & 1.92 \\
Proxima Cen & f1550c & Saturn & Full Clouds & 2 & 1.55 & 3.16 & 3.48 \\
Proxima Cen & f1550c & Saturn & Full Clouds & 5 & 3.88 & 7.91 & 3.95 \\
Proxima Cen & f1550c & Saturn & Patchy Clouds & 1 & 0.78 & 1.58 & 3.11 \\
Proxima Cen & f1550c & Saturn & Patchy Clouds & 2 & 1.55 & 3.16 & 5.62 \\
Proxima Cen & f1550c & Saturn & Patchy Clouds & 5 & 3.88 & 7.91 & 6.37 \\
Proxima Cen & f1550c & Saturn & No Clouds & 1 & 0.78 & 1.58 & 13.76 \\
Proxima Cen & f1550c & Saturn & No Clouds & 2 & 1.55 & 3.16 & 24.86 \\
Proxima Cen & f1550c & Saturn & No Clouds & 5 & 3.88 & 7.91 & 28.16 \\
\enddata
 \label{table:max_snrs}
 \end{deluxetable*}
\tablecomments{Full Exposure Time/SNR tables are provided as Machine Readable Tables.}

\bibliography{biblio}{}

\begin{thebibliography}{}
\expandafter\ifx\csname natexlab\endcsname\relax\def\natexlab#1{#1}\fi
\providecommand{\url}[1]{\href{#1}{#1}}

\bibitem[{{Akeson} {et~al.}(2013){Akeson}, {Chen}, {Ciardi}, {Crane}, {Good},
  {Harbut}, {Jackson}, {Kane}, {Laity}, {Leifer}, {Lynn}, {McElroy}, {Papin},
  {Plavchan}, {Ram{\'{\i}}rez}, {Rey}, {von Braun}, {Wittman}, {Abajian},
  {Ali}, {Beichman}, {Beekley}, {Berriman}, {Berukoff}, {Bryden}, {Chan},
  {Groom}, {Lau}, {Payne}, {Regelson}, {Saucedo}, {Schmitz}, {Stauffer},
  {Wyatt}, \& {Zhang}}]{ExoplanetArchive}
{Akeson}, R.~L., {Chen}, X., {Ciardi}, D., {et~al.} 2013, \pasp, 125, 989

\bibitem[{{Apps} {et~al.}(2010){Apps}, {Clubb}, {Fischer}, {Gaidos}, {Howard},
  {Johnson}, {Marcy}, {Isaacson}, {Giguere}, {Valenti}, {Rodriguez}, {Chubak},
  \& {Lepine}}]{apps2010}
{Apps}, K., {Clubb}, K.~I., {Fischer}, D.~A., {et~al.} 2010, \pasp, 122, 156

\bibitem[{{Astropy Collaboration} {et~al.}(2013){Astropy Collaboration},
  {Robitaille}, {Tollerud}, {Greenfield}, {Droettboom}, {Bray}, {Aldcroft},
  {Davis}, {Ginsburg}, {Price-Whelan}, {Kerzendorf}, {Conley}, {Crighton},
  {Barbary}, {Muna}, {Ferguson}, {Grollier}, {Parikh}, {Nair}, {Unther},
  {Deil}, {Woillez}, {Conseil}, {Kramer}, {Turner}, {Singer}, {Fox}, {Weaver},
  {Zabalza}, {Edwards}, {Azalee Bostroem}, {Burke}, {Casey}, {Crawford},
  {Dencheva}, {Ely}, {Jenness}, {Labrie}, {Lim}, {Pierfederici}, {Pontzen},
  {Ptak}, {Refsdal}, {Servillat}, \& {Streicher}}]{astropy13}
{Astropy Collaboration}, {Robitaille}, T.~P., {Tollerud}, E.~J., {et~al.} 2013,
  aap, 558, A33

\bibitem[{{Astropy Collaboration} {et~al.}(2018){Astropy Collaboration},
  {Price-Whelan}, {Sip{H o}cz}, {G{"u}nther}, {Lim}, {Crawford}, {Conseil},
  {Shupe}, {Craig}, {Dencheva}, {Ginsburg}, {VanderPlas}, {Bradley},
  {P{'e}rez-Su{'a}rez}, {de Val-Borro}, {Aldcroft}, {Cruz}, {Robitaille},
  {Tollerud}, {Ardelean}, {Babej}, {Bach}, {Bachetti}, {Bakanov}, {Bamford},
  {Barentsen}, {Barmby}, {Baumbach}, {Berry}, {Biscani}, {Boquien}, {Bostroem},
  {Bouma}, {Brammer}, {Bray}, {Breytenbach}, {Buddelmeijer}, {Burke},
  {Calderone}, {Cano Rodr{'{i}}guez}, {Cara}, {Cardoso}, {Cheedella}, {Copin},
  {Corrales}, {Crichton}, {D'Avella}, {Deil}, {Depagne}, {Dietrich}, {Donath},
  {Droettboom}, {Earl}, {Erben}, {Fabbro}, {Ferreira}, {Finethy}, {Fox},
  {Garrison}, {Gibbons}, {Goldstein}, {Gommers}, {Greco}, {Greenfield},
  {Groener}, {Grollier}, {Hagen}, {Hirst}, {Homeier}, {Horton}, {Hosseinzadeh},
  {Hu}, {Hunkeler}, {Ivezi{'c}}, {Jain}, {Jenness}, {Kanarek}, {Kendrew},
  {Kern}, {Kerzendorf}, {Khvalko}, {King}, {Kirkby}, {Kulkarni}, {Kumar},
  {Lee}, {Lenz}, {Littlefair}, {Ma}, {Macleod}, {Mastropietro}, {McCully},
  {Montagnac}, {Morris}, {Mueller}, {Mumford}, {Muna}, {Murphy}, {Nelson},
  {Nguyen}, {Ninan}, {N{"o}the}, {Ogaz}, {Oh}, {Parejko}, {Parley}, {Pascual},
  {Patil}, {Patil}, {Plunkett}, {Prochaska}, {Rastogi}, {Reddy Janga},
  {Sabater}, {Sakurikar}, {Seifert}, {Sherbert}, {Sherwood-Taylor}, {Shih},
  {Sick}, {Silbiger}, {Singanamalla}, {Singer}, {Sladen}, {Sooley},
  {Sornarajah}, {Streicher}, {Teuben}, {Thomas}, {Tremblay}, {Turner},
  {Terr{'o}n}, {van Kerkwijk}, {de la Vega}, {Watkins}, {Weaver}, {Whitmore},
  {Woillez}, {Zabalza}, \& {Astropy Contributors}}]{astropy18}
{Astropy Collaboration}, {Price-Whelan}, A.~M., {Sip{H o}cz}, B.~M., {et~al.}
  2018, aj, 156, 123

\bibitem[{{Bailey} {et~al.}(2009){Bailey}, {Butler}, {Tinney}, {Jones},
  {O'Toole}, {Carter}, \& {Marcy}}]{bailey2009}
{Bailey}, J., {Butler}, R.~P., {Tinney}, C.~G., {et~al.} 2009, \apj, 690, 743

\bibitem[{{Bakos} {et~al.}(2018){Bakos}, {Bayliss}, {Bento}, {Bhatti}, {Brahm},
  {Csubry}, {Espinoza}, {Hartman}, {Henning}, {Jord{\'a}n}, {Mancini}, {Penev},
  {Rabus}, {Sarkis}, {Suc}, {de Val-Borro}, {Zhou}, {Butler}, {Crane},
  {Durkan}, {Shectman}, {Kim}, {L{\'a}z{\'a}r}, {Papp}, {S{\'a}ri}, {Ricker},
  {Vanderspek}, {Latham}, {Seager}, {Winn}, {Jenkins}, {Chacon},
  {F{\'{u}}r{\'e}sz}, {Goeke}, {Li}, {Quinn}, {Quintana}, {Tenenbaum}, {Teske},
  {Vezie}, {Yu}, {Stockdale}, {Evans}, \& {Relles}}]{bakos2018}
{Bakos}, G.~{\'A}., {Bayliss}, D., {Bento}, J., {et~al.} 2018, arXiv e-prints,
  arXiv:1812.09406

\bibitem[{{Bayliss} {et~al.}(2018){Bayliss}, {Gillen}, {Eigm{\"u}ller},
  {McCormac}, {Alexander}, {Armstrong}, {Booth}, {Bouchy}, {Burleigh},
  {Cabrera}, {Casewell}, {Chaushev}, {Chazelas}, {Csizmadia}, {Erikson},
  {Faedi}, {Foxell}, {G{\"a}nsicke}, {Goad}, {Grange}, {G{\"u}nther},
  {Hodgkin}, {Jackman}, {Jenkins}, {Lambert}, {Louden}, {Metrailler}, {Moyano},
  {Pollacco}, {Poppenhaeger}, {Queloz}, {Raddi}, {Rauer}, {Raynard}, {Smith},
  {Soto}, {Thompson}, {Titz-Weider}, {Udry}, {Walker}, {Watson}, {West}, \&
  {Wheatley}}]{bayliss2018}
{Bayliss}, D., {Gillen}, E., {Eigm{\"u}ller}, P., {et~al.} 2018, \mnras, 475,
  4467

\bibitem[{{Beichman} {et~al.}(2010){Beichman}, {Krist}, {Trauger}, {Greene},
  {Oppenheimer}, {Sivaramakrishnan}, {Doyon}, {Boccaletti}, {Barman}, \&
  {Rieke}}]{beichman2010}
{Beichman}, C.~A., {Krist}, J., {Trauger}, J.~T., {et~al.} 2010, Publications
  of the Astronomical Society of the Pacific, 122, 162

\bibitem[{{Berger} {et~al.}(2006){Berger}, {Gies}, {McAlister}, {ten
  Brummelaar}, {Henry}, {Sturmann}, {Sturmann}, {Turner}, {Ridgway},
  {Aufdenberg}, \& {M{\'e}rand}}]{berger2006}
{Berger}, D.~H., {Gies}, D.~R., {McAlister}, H.~A., {et~al.} 2006, \apj, 644,
  475

\bibitem[{{Bobylev}(2010)}]{bobylev2010}
{Bobylev}, V.~V. 2010, Astronomy Letters, 36, 220

\bibitem[{{Boccaletti} {et~al.}(2015){Boccaletti}, {Lagage}, {Baudoz},
  {Beichman}, {Bouchet}, {Cavarroc}, {Dubreuil}, {Glasse}, {Glauser}, {Hines},
  {Lajoie}, {Lebreton}, {Perrin}, {Pueyo}, {Reess}, {Rieke}, {Ronayette},
  {Rouan}, {Soummer}, \& {Wright}}]{boccaletti2015}
{Boccaletti}, A., {Lagage}, P.~O., {Baudoz}, P., {et~al.} 2015, Publications of
  the Astronomical Society of the Pacific, 127, 633

\bibitem[{{Bouchet} {et~al.}(2015){Bouchet}, {Garc{\'\i}a-Mar{\'\i}n},
  {Lagage}, {Amiaux}, {Augu{\'e}res}, {Bauwens}, {Blommaert}, {Chen}, {Detre},
  {Dicken}, {Dubreuil}, {Galdemard}, {Gastaud}, {Glasse}, {Gordon}, {Gougnaud},
  {Guillard}, {Justtanont}, {Krause}, {Leboeuf}, {Longval}, {Martin}, {Mazy},
  {Moreau}, {Olofsson}, {Ray}, {Rees}, {Renotte}, {Ressler}, {Ronayette},
  {Salasca}, {Scheithauer}, {Sykes}, {Thelen}, {Wells}, {Wright}, \&
  {Wright}}]{Bouchet2015}
{Bouchet}, P., {Garc{\'\i}a-Mar{\'\i}n}, M., {Lagage}, P.~O., {et~al.} 2015,
  Publications of the Astronomical Society of the Pacific, 127, 612

\bibitem[{{Bowler}(2016)}]{bowler2016}
{Bowler}, B.~P. 2016, Publications of the Astronomical Society of the Pacific,
  128, 102001

\bibitem[{{Brandt} {et~al.}(2019){Brandt}, {Dupuy}, \& {Bowler}}]{brandt2019}
{Brandt}, T.~D., {Dupuy}, T.~J., \& {Bowler}, B.~P. 2019, \aj, 158, 140

\bibitem[{{Bryan} {et~al.}(2016){Bryan}, {Knutson}, {Howard}, {Ngo}, {Batygin},
  {Crepp}, {Fulton}, {Hinkley}, {Isaacson}, {Johnson}, {Marcy}, \&
  {Wright}}]{bryan2016}
{Bryan}, M.~L., {Knutson}, H.~A., {Howard}, A.~W., {et~al.} 2016, \apj, 821, 89

\bibitem[{{Burke} {et~al.}(2015){Burke}, {Christiansen}, {Mullally}, {Seader},
  {Huber}, {Rowe}, {Coughlin}, {Thompson}, {Catanzarite}, {Clarke}, {Morton},
  {Caldwell}, {Bryson}, {Haas}, {Batalha}, {Jenkins}, {Tenenbaum}, {Twicken},
  {Li}, {Quintana}, {Barclay}, {Henze}, {Borucki}, {Howell}, \&
  {Still}}]{Burke2015}
{Burke}, C.~J., {Christiansen}, J.~L., {Mullally}, F., {et~al.} 2015, \apj,
  809, 8

\bibitem[{{Butler} {et~al.}(2006){Butler}, {Johnson}, {Marcy}, {Wright},
  {Vogt}, \& {Fischer}}]{butler2006}
{Butler}, R.~P., {Johnson}, J.~A., {Marcy}, G.~W., {et~al.} 2006, \pasp, 118,
  1685

\bibitem[{{Cassan} {et~al.}(2012){Cassan}, {Kubas}, {Beaulieu}, {Dominik},
  {Horne}, {Greenhill}, {Wambsganss}, {Menzies}, {Williams}, {J{\o}rgensen},
  {Udalski}, {Bennett}, {Albrow}, {Batista}, {Brillant}, {Caldwell}, {Cole},
  {Coutures}, {Cook}, {Dieters}, {Dominis Prester}, {Donatowicz}, {Fouqu{\'e}},
  {Hill}, {Kains}, {Kane}, {Marquette}, {Martin}, {Pollard}, {Sahu}, {Vinter},
  {Warren}, {Watson}, {Zub}, {Sumi}, {Szyma{\'n}ski}, {Kubiak}, {Poleski},
  {Soszynski}, {Ulaczyk}, {Pietrzy{\'n}ski}, \& {Wyrzykowski}}]{cassan2012}
{Cassan}, A., {Kubas}, D., {Beaulieu}, J.~P., {et~al.} 2012, \nat, 481, 167

\bibitem[{{Chauvin} {et~al.}(2017){Chauvin}, {Desidera}, {Lagrange}, {Vigan},
  {Gratton}, {Langlois}, {Bonnefoy}, {Beuzit}, {Feldt}, {Mouillet}, {Meyer},
  {Cheetham}, {Biller}, {Boccaletti}, {D'Orazi}, {Galicher}, {Hagelberg},
  {Maire}, {Mesa}, {Olofsson}, {Samland}, {Schmidt}, {Sissa}, {Bonavita},
  {Charnay}, {Cudel}, {Daemgen}, {Delorme}, {Janin-Potiron}, {Janson},
  {Keppler}, {Le Coroller}, {Ligi}, {Marleau}, {Messina}, {Molli{\`e}re},
  {Mordasini}, {M{\"u}ller}, {Peretti}, {Perrot}, {Rodet}, {Rouan}, {Zurlo},
  {Dominik}, {Henning}, {Menard}, {Schmid}, {Turatto}, {Udry}, {Vakili}, {Abe},
  {Antichi}, {Baruffolo}, {Baudoz}, {Baudrand}, {Blanchard}, {Bazzon}, {Buey},
  {Carbillet}, {Carle}, {Charton}, {Cascone}, {Claudi}, {Costille}, {Deboulbe},
  {De Caprio}, {Dohlen}, {Fantinel}, {Feautrier}, {Fusco}, {Gigan}, {Giro},
  {Gisler}, {Gluck}, {Hubin}, {Hugot}, {Jaquet}, {Kasper}, {Madec}, {Magnard},
  {Martinez}, {Maurel}, {Le Mignant}, {M{\"o}ller-Nilsson}, {Llored}, {Moulin},
  {Orign{\'e}}, {Pavlov}, {Perret}, {Petit}, {Pragt}, {Puget}, {Rabou},
  {Ramos}, {Rigal}, {Rochat}, {Roelfsema}, {Rousset}, {Roux}, {Salasnich},
  {Sauvage}, {Sevin}, {Soenke}, {Stadler}, {Suarez}, {Weber}, {Wildi},
  {Antoniucci}, {Augereau}, {Baudino}, {Brandner}, {Engler}, {Girard}, {Gry},
  {Kral}, {Kopytova}, {Lagadec}, {Milli}, {Moutou}, {Schlieder},
  {Szul{\'a}gyi}, {Thalmann}, \& {Wahhaj}}]{chauvin2017}
{Chauvin}, G., {Desidera}, S., {Lagrange}, A.~M., {et~al.} 2017, \aap, 605, L9

\bibitem[{{Chen} \& {Kipping}(2017)}]{forecaster}
{Chen}, J., \& {Kipping}, D. 2017, {Forecaster: Mass and radii of planets
  predictor}, , , ascl:1701.007

\bibitem[{{Clanton} \& {Gaudi}(2016)}]{clanton2016}
{Clanton}, C., \& {Gaudi}, B.~S. 2016, \apj, 819, 125

\bibitem[{{Davison} {et~al.}(2015){Davison}, {White}, {Henry}, {Riedel}, {Jao},
  {Bailey}, {Quinn}, {Cantrell}, {Subasavage}, \& {Winters}}]{davison2015}
{Davison}, C.~L., {White}, R.~J., {Henry}, T.~J., {et~al.} 2015, \aj, 149, 106

\bibitem[{{Dawson} \& {De Robertis}(2004)}]{dawson2004}
{Dawson}, P.~C., \& {De Robertis}, M.~M. 2004, \aj, 127, 2909

\bibitem[{{Demory} {et~al.}(2009){Demory}, {S{\'e}gransan}, {Forveille},
  {Queloz}, {Beuzit}, {Delfosse}, {di Folco}, {Kervella}, {Le Bouquin},
  {Perrier}, {Benisty}, {Duvert}, {Hofmann}, {Lopez}, \& {Petrov}}]{demory2009}
{Demory}, B.~O., {S{\'e}gransan}, D., {Forveille}, T., {et~al.} 2009, \aap,
  505, 205

\bibitem[{{Dieterich} {et~al.}(2014){Dieterich}, {Henry}, {Jao}, {Winters},
  {Hosey}, {Riedel}, \& {Subasavage}}]{dieterich2014}
{Dieterich}, S.~B., {Henry}, T.~J., {Jao}, W.-C., {et~al.} 2014, \aj, 147, 94

\bibitem[{{Doyle} \& {Butler}(1990)}]{doyle1990}
{Doyle}, J.~G., \& {Butler}, C.~J. 1990, \aap, 235, 335

\bibitem[{Foreman-Mackey(2018)}]{exoplanet}
Foreman-Mackey, D. 2018, exoplanet v0.1.3, , , doi:10.5281/zenodo.2536576.
\newblock \url{https://doi.org/10.5281/zenodo.2536576}

\bibitem[{{Forveille} {et~al.}(2011){Forveille}, {Bonfils}, {Lo Curto},
  {Delfosse}, {Udry}, {Bouchy}, {Lovis}, {Mayor}, {Moutou}, {Naef}, {Pepe},
  {Perrier}, {Queloz}, \& {Santos}}]{forveille2011}
{Forveille}, T., {Bonfils}, X., {Lo Curto}, G., {et~al.} 2011, \aap, 526, A141

\bibitem[{{Girard} {et~al.}(2018){Girard}, {Blair}, {Brooks}, {Brooks},
  {Brown}, {Bushouse}, {Canipe}, {Chen}, {Correnti}, {Hagan}, {Hilbert},
  {Hines}, {Leisenring}, {Long}, {Nickson}, {Perrin}, {Pontoppidan}, {Pueyo},
  {Rajan}, {Riedel}, {Soummer}, {Stansberry}, {Stark}, {Van Gorkom}, \&
  {York}}]{girard2018}
{Girard}, J.~H., {Blair}, W., {Brooks}, B., {et~al.} 2018, in Society of
  Photo-Optical Instrumentation Engineers (SPIE) Conference Series, Vol. 10698,
  \procspie, 106983V

\bibitem[{{Haghighipour} {et~al.}(2010){Haghighipour}, {Vogt}, {Butler},
  {Rivera}, {Laughlin}, {Meschiari}, \& {Henry}}]{haghighipour2010}
{Haghighipour}, N., {Vogt}, S.~S., {Butler}, R.~P., {et~al.} 2010, \apj, 715,
  271

\bibitem[{{Hartman} {et~al.}(2015){Hartman}, {Bayliss}, {Brahm}, {Bakos},
  {Mancini}, {Jord{\'a}n}, {Penev}, {Rabus}, {Zhou}, {Butler}, {Espinoza}, {de
  Val-Borro}, {Bhatti}, {Csubry}, {Ciceri}, {Henning}, {Schmidt}, {Arriagada},
  {Shectman}, {Crane}, {Thompson}, {Suc}, {Cs{\'a}k}, {Tan}, {Noyes},
  {L{\'a}z{\'a}r}, {Papp}, \& {S{\'a}ri}}]{hartman2015}
{Hartman}, J.~D., {Bayliss}, D., {Brahm}, R., {et~al.} 2015, \aj, 149, 166

\bibitem[{{Henry} {et~al.}(2006){Henry}, {Jao}, {Subasavage}, {Beaulieu},
  {Ianna}, {Costa}, \& {M{\'e}ndez}}]{henry2006}
{Henry}, T.~J., {Jao}, W.-C., {Subasavage}, J.~P., {et~al.} 2006, \aj, 132,
  2360

\bibitem[{{Howard} {et~al.}(2010){Howard}, {Johnson}, {Marcy}, {Fischer},
  {Wright}, {Bernat}, {Henry}, {Peek}, {Isaacson}, {Apps}, {Endl}, {Cochran},
  {Valenti}, {Anderson}, \& {Piskunov}}]{howard2010}
{Howard}, A.~W., {Johnson}, J.~A., {Marcy}, G.~W., {et~al.} 2010, \apj, 721,
  1467

\bibitem[{Hunter(2007)}]{matplotlib}
Hunter, J.~D. 2007, Computing In Science \& Engineering, 9, 90

\bibitem[{{Johnson} \& {Wright}(1983)}]{johnson1983}
{Johnson}, H.~M., \& {Wright}, C.~D. 1983, The Astrophysical Journal Supplement
  Series, 53, 643

\bibitem[{{Johnson} {et~al.}(2007){Johnson}, {Butler}, {Marcy}, {Fischer},
  {Vogt}, {Wright}, \& {Peek}}]{johnson2007}
{Johnson}, J.~A., {Butler}, R.~P., {Marcy}, G.~W., {et~al.} 2007, \apj, 670,
  833

\bibitem[{{Johnson} {et~al.}(2010){Johnson}, {Howard}, {Marcy}, {Bowler},
  {Henry}, {Fischer}, {Apps}, {Isaacson}, \& {Wright}}]{johnson2010}
{Johnson}, J.~A., {Howard}, A.~W., {Marcy}, G.~W., {et~al.} 2010, \pasp, 122,
  149

\bibitem[{{Johnson} {et~al.}(2012){Johnson}, {Gazak}, {Apps}, {Muirhead},
  {Crepp}, {Crossfield}, {Boyajian}, {von Braun}, {Rojas-Ayala}, {Howard},
  {Covey}, {Schlawin}, {Hamren}, {Morton}, {Marcy}, \& {Lloyd}}]{johnson2012}
{Johnson}, J.~A., {Gazak}, J.~Z., {Apps}, K., {et~al.} 2012, \aj, 143, 111

\bibitem[{{Kalas} {et~al.}(2008){Kalas}, {Graham}, {Chiang}, {Fitzgerald},
  {Clampin}, {Kite}, {Stapelfeldt}, {Marois}, \& {Krist}}]{kalas2008}
{Kalas}, P., {Graham}, J.~R., {Chiang}, E., {et~al.} 2008, Science, 322, 1345

\bibitem[{{Keppler} {et~al.}(2018){Keppler}, {Benisty}, {M{\"u}ller},
  {Henning}, {van Boekel}, {Cantalloube}, {Ginski}, {van Holstein}, {Maire},
  {Pohl}, {Samland }, {Avenhaus}, {Baudino}, {Boccaletti}, {de Boer},
  {Bonnefoy}, {Chauvin}, {Desidera}, {Langlois}, {Lazzoni}, {Marleau},
  {Mordasini}, {Pawellek}, {Stolker}, {Vigan}, {Zurlo}, {Birnstiel},
  {Brandner}, {Feldt}, {Flock}, {Girard}, {Gratton}, {Hagelberg}, {Isella},
  {Janson}, {Juhasz}, {Kemmer}, {Kral}, {Lagrange}, {Launhardt}, {Matter},
  {M{\'e}nard}, {Milli}, {Molli{\`e}re}, {Olofsson}, {P{\'e}rez}, {Pinilla},
  {Pinte}, {Quanz}, {Schmidt}, {Udry}, {Wahhaj}, {Williams}, {Buenzli},
  {Cudel}, {Dominik}, {Galicher}, {Kasper}, {Lannier}, {Mesa}, {Mouillet},
  {Peretti}, {Perrot}, {Salter}, {Sissa}, {Wildi}, {Abe}, {Antichi},
  {Augereau}, {Baruffolo}, {Baudoz}, {Bazzon}, {Beuzit}, {Blanchard}, {Brems},
  {Buey}, {De Caprio}, {Carbillet}, {Carle}, {Cascone}, {Cheetham}, {Claudi},
  {Costille}, {Delboulb{\'e}}, {Dohlen}, {Fantinel}, {Feautrier}, {Fusco},
  {Giro}, {Gluck}, {Gry}, {Hubin}, {Hugot}, {Jaquet}, {Le Mignant}, {Llored},
  {Madec}, {Magnard}, {Martinez}, {Maurel}, {Meyer}, {M{\"o}ller-Nilsson},
  {Moulin}, {Mugnier}, {Orign{\'e}}, {Pavlov}, {Perret}, {Petit}, {Pragt},
  {Puget}, {Rabou}, {Ramos}, {Rigal}, {Rochat}, {Roelfsema}, {Rousset}, {Roux},
  {Salasnich}, {Sauvage}, {Sevin}, {Soenke}, {Stadler}, {Suarez}, {Turatto}, \&
  {Weber}}]{keppler2018}
{Keppler}, M., {Benisty}, M., {M{\"u}ller}, A., {et~al.} 2018, \aap, 617, A44

\bibitem[{{Lacy}(1977)}]{lacy1977}
{Lacy}, C.~H. 1977, The Astrophysical Journal Supplement Series, 34, 479

\bibitem[{{Lagrange} {et~al.}(2009){Lagrange}, {Gratadour}, {Chauvin}, {Fusco},
  {Ehrenreich}, {Mouillet}, {Rousset}, {Rouan}, {Allard}, {Gendron}, {Charton},
  {Mugnier}, {Rabou}, {Montri}, \& {Lacombe}}]{lagrange2009}
{Lagrange}, A.~M., {Gratadour}, D., {Chauvin}, G., {et~al.} 2009, \aap, 493,
  L21

\bibitem[{{Lajoie} {et~al.}(2016){Lajoie}, {Soummer}, {Pueyo}, {Hines},
  {Nelan}, {Perrin}, {Clampin}, \& {Isaacs}}]{lajoie2016}
{Lajoie}, C.-P., {Soummer}, R., {Pueyo}, L., {et~al.} 2016, in Society of
  Photo-Optical Instrumentation Engineers (SPIE) Conference Series, Vol. 9904,
  \procspie, 99045K

\bibitem[{{Lannier} {et~al.}(2016){Lannier}, {Delorme}, {Lagrange}, {Borgniet},
  {Rameau}, {Schlieder}, {Gagn{\'e}}, {Bonavita}, {Malo}, {Chauvin},
  {Bonnefoy}, \& {Girard}}]{lannier2016}
{Lannier}, J., {Delorme}, P., {Lagrange}, A.~M., {et~al.} 2016, \aap, 596, A83

\bibitem[{{Linsky} {et~al.}(1995){Linsky}, {Wood}, {Brown}, {Giampapa}, \&
  {Ambruster}}]{linsky1995}
{Linsky}, J.~L., {Wood}, B.~E., {Brown}, A., {Giampapa}, M.~S., \& {Ambruster},
  C. 1995, \apj, 455, 670

\bibitem[{{Luhman} {et~al.}(2011){Luhman}, {Burgasser}, \&
  {Bochanski}}]{luhman2011}
{Luhman}, K.~L., {Burgasser}, A.~J., \& {Bochanski}, J.~J. 2011, \apj, 730, L9

\bibitem[{{Luhman} {et~al.}(2009){Luhman}, {Mamajek}, {Allen}, {Muench}, \&
  {Finkbeiner}}]{luhman2009}
{Luhman}, K.~L., {Mamajek}, E.~E., {Allen}, P.~R., {Muench}, A.~A., \&
  {Finkbeiner}, D.~P. 2009, \apj, 691, 1265

\bibitem[{{Luhman} {et~al.}(2006){Luhman}, {Wilson}, {Brandner}, {Skrutskie},
  {Nelson}, {Smith}, {Peterson}, {Cushing}, \& {Young}}]{luhman2006}
{Luhman}, K.~L., {Wilson}, J.~C., {Brandner}, W., {et~al.} 2006, \apj, 649, 894

\bibitem[{{Luhman} {et~al.}(2007){Luhman}, {Patten}, {Marengo}, {Schuster},
  {Hora}, {Ellis}, {Stauffer}, {Sonnett}, {Winston}, {Gutermuth}, {Megeath},
  {Backman}, {Henry}, {Werner}, \& {Fazio}}]{luhman2007}
{Luhman}, K.~L., {Patten}, B.~M., {Marengo}, M., {et~al.} 2007, \apj, 654, 570

\bibitem[{{Macintosh} {et~al.}(2015){Macintosh}, {Graham}, {Barman}, {De Rosa},
  {Konopacky}, {Marley}, {Marois}, {Nielsen}, {Pueyo}, {Rajan}, {Rameau},
  {Saumon}, {Wang}, {Patience}, {Ammons}, {Arriaga}, {Artigau}, {Beckwith},
  {Brewster}, {Bruzzone}, {Bulger}, {Burningham}, {Burrows}, {Chen}, {Chiang},
  {Chilcote}, {Dawson}, {Dong}, {Doyon}, {Draper}, {Duch{\^e}ne}, {Esposito},
  {Fabrycky}, {Fitzgerald}, {Follette}, {Fortney}, {Gerard}, {Goodsell},
  {Greenbaum}, {Hibon}, {Hinkley}, {Cotten}, {Hung}, {Ingraham},
  {Johnson-Groh}, {Kalas}, {Lafreniere}, {Larkin}, {Lee}, {Line}, {Long},
  {Maire}, {Marchis}, {Matthews}, {Max}, {Metchev}, {Millar-Blanchaer},
  {Mittal}, {Morley}, {Morzinski}, {Murray-Clay}, {Oppenheimer}, {Palmer},
  {Patel}, {Perrin}, {Poyneer}, {Rafikov}, {Rantakyr{\"o}}, {Rice}, {Rojo},
  {Rudy}, {Ruffio}, {Ruiz}, {Sadakuni}, {Saddlemyer}, {Salama}, {Savransky},
  {Schneider}, {Sivaramakrishnan}, {Song}, {Soummer}, {Thomas}, {Vasisht},
  {Wallace}, {Ward- Duong}, {Wiktorowicz}, {Wolff}, \&
  {Zuckerman}}]{macintosh2015}
{Macintosh}, B., {Graham}, J.~R., {Barman}, T., {et~al.} 2015, Science, 350, 64

\bibitem[{{Marois} {et~al.}(2008){Marois}, {Macintosh}, {Barman}, {Zuckerman},
  {Song}, {Patience}, {Lafreni{\`e}re}, \& {Doyon}}]{marois2008}
{Marois}, C., {Macintosh}, B., {Barman}, T., {et~al.} 2008, Science, 322, 1348

\bibitem[{{Marois} {et~al.}(2010){Marois}, {Zuckerman}, {Konopacky},
  {Macintosh}, \& {Barman}}]{marois2010}
{Marois}, C., {Zuckerman}, B., {Konopacky}, Q.~M., {Macintosh}, B., \&
  {Barman}, T. 2010, \nat, 468, 1080

\bibitem[{{Mennesson} {et~al.}(2018){Mennesson}, {Debes}, {Douglas}, {Nemati},
  {Stark}, {Kasdin}, {Macintosh}, {Turnbull}, {Rizzo}, {Roberge}, {Zimmerman},
  {Cahoy}, {Krist}, {Bailey}, {Trauger}, {Rhodes}, {Moustakas}, {Frerking},
  {Zhao}, {Poberezhskiy}, \& {Demers}}]{mennesson2018}
{Mennesson}, B., {Debes}, J., {Douglas}, E., {et~al.} 2018, in Society of
  Photo-Optical Instrumentation Engineers (SPIE) Conference Series, Vol. 10698,
  106982I

\bibitem[{{Meyer} {et~al.}(2018){Meyer}, {Amara}, {Reggiani}, \&
  {Quanz}}]{meyer2018}
{Meyer}, M.~R., {Amara}, A., {Reggiani}, M., \& {Quanz}, S.~P. 2018, \aap, 612,
  L3

\bibitem[{{Montet} {et~al.}(2014){Montet}, {Crepp}, {Johnson}, {Howard}, \&
  {Marcy}}]{montet2014}
{Montet}, B.~T., {Crepp}, J.~R., {Johnson}, J.~A., {Howard}, A.~W., \& {Marcy},
  G.~W. 2014, \apj, 781, 28

\bibitem[{{Morin} {et~al.}(2010){Morin}, {Donati}, {Petit}, {Delfosse},
  {Forveille}, \& {Jardine}}]{morin2010}
{Morin}, J., {Donati}, J.~F., {Petit}, P., {et~al.} 2010, \mnras, 407, 2269

\bibitem[{{Mr{\'o}z} {et~al.}(2017){Mr{\'o}z}, {Udalski}, {Skowron}, {Poleski},
  {Koz{\l}owski}, {Szyma{\'n}ski}, {Soszy{\'n}ski}, {Wyrzykowski},
  {Pietrukowicz}, {Ulaczyk}, {Skowron}, \& {Pawlak}}]{mroz2017}
{Mr{\'o}z}, P., {Udalski}, A., {Skowron}, J., {et~al.} 2017, \nat, 548, 183

\bibitem[{{Mulders} {et~al.}(2015){Mulders}, {Pascucci}, \&
  {Apai}}]{Mulders2015}
{Mulders}, G.~D., {Pascucci}, I., \& {Apai}, D. 2015, \apj, 798, 112

\bibitem[{Oliphant(2006--)}]{numpy}
Oliphant, T. 2006--, {NumPy}: A guide to {NumPy}, USA: Trelgol Publishing, , .
\newblock \url{http://www.numpy.org/}

\bibitem[{{Pasinetti Fracassini} {et~al.}(2001){Pasinetti Fracassini},
  {Pastori}, {Covino}, \& {Pozzi}}]{pasinetti2001}
{Pasinetti Fracassini}, L.~E., {Pastori}, L., {Covino}, S., \& {Pozzi}, A.
  2001, \aap, 367, 521

\bibitem[{{Penny} {et~al.}(2019){Penny}, {Gaudi}, {Kerins}, {Rattenbury},
  {Mao}, {Robin}, \& {Calchi Novati}}]{penny2019}
{Penny}, M.~T., {Gaudi}, B.~S., {Kerins}, E., {et~al.} 2019, \apjs, 241, 3

\bibitem[{{Pontoppidan} {et~al.}(2016){Pontoppidan}, {Pickering}, {Laidler},
  {Gilbert}, {Sontag}, {Slocum}, {Sienkiewicz}, {Hanley}, {Earl}, {Pueyo},
  {Ravindranath}, {Karakla}, {Robberto}, {Noriega-Crespo}, \&
  {Barker}}]{pontoppidan2016}
{Pontoppidan}, K.~M., {Pickering}, T.~E., {Laidler}, V.~G., {et~al.} 2016, in
  Society of Photo-Optical Instrumentation Engineers (SPIE) Conference Series,
  Vol. 9910, Observatory Operations: Strategies, Processes, and Systems VI,
  991016

\bibitem[{{Rameau} {et~al.}(2013{\natexlab{a}}){Rameau}, {Chauvin}, {Lagrange},
  {Boccaletti}, {Quanz}, {Bonnefoy}, {Girard}, {Delorme}, {Desidera}, {Klahr},
  {Mordasini}, {Dumas}, \& {Bonavita}}]{rameau2013a}
{Rameau}, J., {Chauvin}, G., {Lagrange}, A.~M., {et~al.} 2013{\natexlab{a}},
  \apj, 772, L15

\bibitem[{{Rameau} {et~al.}(2013{\natexlab{b}}){Rameau}, {Chauvin}, {Lagrange},
  {Meshkat}, {Boccaletti}, {Quanz}, {Currie}, {Mawet}, {Girard}, {Bonnefoy}, \&
  {Kenworthy}}]{rameau2013b}
---. 2013{\natexlab{b}}, \apj, 779, L26

\bibitem[{{Reiners} {et~al.}(2009){Reiners}, {Basri}, \&
  {Browning}}]{reiners2009}
{Reiners}, A., {Basri}, G., \& {Browning}, M. 2009, \apj, 692, 538

\bibitem[{{Rieke} {et~al.}(2015){Rieke}, {Wright}, {B{\"o}ker}, {Bouwman},
  {Colina}, {Glasse}, {Gordon}, {Greene}, {G{\"u}del}, {Henning}, {Justtanont},
  {Lagage}, {Meixner}, {N{\o}rgaard-Nielsen}, {Ray}, {Ressler}, {van Dishoeck},
  \& {Waelkens}}]{rieke2015}
{Rieke}, G.~H., {Wright}, G.~S., {B{\"o}ker}, T., {et~al.} 2015, Publications
  of the Astronomical Society of the Pacific, 127, 584

\bibitem[{{Robertson} {et~al.}(2013){Robertson}, {Endl}, {Cochran}, {MacQueen},
  \& {Boss}}]{robertson2013}
{Robertson}, P., {Endl}, M., {Cochran}, W.~D., {MacQueen}, P.~J., \& {Boss},
  A.~P. 2013, \apj, 774, 147

\bibitem[{{Sahlmann} {et~al.}(2016){Sahlmann}, {Lazorenko}, {S{\'e}gransan},
  {Astudillo-Defru}, {Bonfils}, {Delfosse}, {Forveille}, {Hagelberg}, {Lo
  Curto}, {Pepe}, {Queloz}, {Udry}, \& {Zimmerman}}]{sahlmann2016}
{Sahlmann}, J., {Lazorenko}, P.~F., {S{\'e}gransan}, D., {et~al.} 2016, \aap,
  595, A77

\bibitem[{Salvatier {et~al.}(2016)Salvatier, Wiecki, \& Fonnesbeck}]{PyMC3}
Salvatier, J., Wiecki, T.~V., \& Fonnesbeck, C. 2016, PeerJ Computer Science,
  2, e55

\bibitem[{{Schlieder} {et~al.}(2015){Schlieder}, {Beichman}, {Meyer}, \&
  {Greene}}]{schlieder2015}
{Schlieder}, J.~E., {Beichman}, C.~A., {Meyer}, M.~R., \& {Greene}, T. 2015,
  arXiv e-prints, arXiv:1512.00053

\bibitem[{{Shvartzvald} {et~al.}(2016){Shvartzvald}, {Maoz}, {Udalski}, {Sumi},
  {Friedmann}, {Kaspi}, {Poleski}, {Szyma{\'n}ski}, {Skowron}, {Koz{\l}owski},
  {Wyrzykowski}, {Mr{\'o}z}, {Pietrukowicz}, {Pietrzy{\'n}ski},
  {Soszy{\'n}ski}, {Ulaczyk}, {Abe}, {Barry}, {Bennett}, {Bhattacharya},
  {Bond}, {Freeman}, {Inayama}, {Itow}, {Koshimoto}, {Ling}, {Masuda}, {Fukui},
  {Matsubara}, {Muraki}, {Ohnishi}, {Rattenbury}, {Saito}, {Sullivan},
  {Suzuki}, {Tristram}, {Wakiyama}, \& {Yonehara}}]{shvartzvald2016}
{Shvartzvald}, Y., {Maoz}, D., {Udalski}, A., {et~al.} 2016, \mnras, 457, 4089

\bibitem[{STScI(2016)}]{jdox2016}
STScI. 2016, "JWST User Documentation",  Baltimore, MD: Space Telescope Science
  Institute.
\newblock \url{"https://jwst-docs.stsci.edu/"}

\bibitem[{STScI(2017{\natexlab{a}})}]{jdoxetccoro2016}
---. 2017{\natexlab{a}}, "JWST Coronagraphy in ETC",  Baltimore, MD: Space
  Telescope Science Institute.
\newblock
  \url{"https://jwst-docs.stsci.edu/methods-and-roadmaps/jwst-high-contrast-imaging/jwst-coronagraphy-in-etc"}

\bibitem[{STScI(2017{\natexlab{b}})}]{etccorostrat2016}
---. 2017{\natexlab{b}}, "JWST ETC Coronagraphy Strategy",  Baltimore, MD:
  Space Telescope Science Institute.
\newblock
  \url{"https://jwst-docs.stsci.edu/jwst-exposure-time-calculator-overview/jwst-etc-calculations-page-overview/jwst-etc-strategies/jwst-etc-coronagraphy-strategy"}

\bibitem[{{Suzuki} {et~al.}(2016){Suzuki}, {Bennett}, {Sumi}, {Bond}, {Rogers},
  {Abe}, {Asakura}, {Bhattacharya}, {Donachie}, {Freeman}, {Fukui}, {Hirao},
  {Itow}, {Koshimoto}, {Li}, {Ling}, {Masuda}, {Matsubara}, {Muraki},
  {Nagakane}, {Onishi}, {Oyokawa}, {Rattenbury}, {Saito}, {Sharan}, {Shibai},
  {Sullivan}, {Tristram}, {Yonehara}, \& {MOA Collaboration}}]{suzuki2016}
{Suzuki}, D., {Bennett}, D.~P., {Sumi}, T., {et~al.} 2016, \apj, 833, 145

\bibitem[{{The HabEx Study Team}(2019)}]{habex2019}
{The HabEx Study Team}. 2019, {The Habitable Exoplanet Observatory Final
  Report}, , .
\newblock
  \url{"https://www.jpl.nasa.gov/habex/pdf/HabEx-Final-Report-Public-Release-LINKED-0924.pdf"}

\bibitem[{{The LUVOIR Study Team}(2019)}]{luvoir2019}
{The LUVOIR Study Team}. 2019, {The Large UV Optical Infrared Surveyor Final
  Report}, , .
\newblock
  \url{"https://asd.gsfc.nasa.gov/luvoir/reports/LUVOIR_FinalReport_2019-08-26.pdf"}

\bibitem[{{Theano Development Team}(2016)}]{theano}
{Theano Development Team}. 2016, arXiv e-prints, abs/1605.02688.
\newblock \url{http://arxiv.org/abs/1605.02688}

\bibitem[{{Todorov} {et~al.}(2010){Todorov}, {Luhman}, \&
  {McLeod}}]{todorov2010}
{Todorov}, K., {Luhman}, K.~L., \& {McLeod}, K.~K. 2010, \apj, 714, L84

\bibitem[{{Trifonov} {et~al.}(2018){Trifonov}, {K{\"u}rster}, {Zechmeister},
  {Tal-Or}, {Caballero}, {Quirrenbach}, {Amado}, {Ribas}, {Reiners}, {Reffert},
  {Dreizler}, {Hatzes}, {Kaminski}, {Launhardt}, {Henning}, {Montes},
  {B{\'e}jar}, {Mundt}, {Pavlov}, {Schmitt}, {Seifert}, {Morales}, {Nowak},
  {Jeffers}, {Rodr{\'{\i}}guez-L{\'o}pez}, {del Burgo}, {Anglada-Escud{\'e}},
  {L{\'o}pez-Santiago}, {Mathar}, {Ammler-von Eiff}, {Guenther}, {Barrado},
  {Gonz{\'a}lez Hern{\'a}ndez}, {Mancini}, {St{\"u}rmer}, {Abril}, {Aceituno},
  {Alonso-Floriano}, {Antona}, {Anwand-Heerwart}, {Arroyo-Torres}, {Azzaro},
  {Baroch}, {Bauer}, {Becerril}, {Ben{\'{\i}}tez}, {Berdi{\~n}as}, {Bergond},
  {Bl{\"u}mcke}, {Brinkm{\"o}ller}, {Cano}, {C{\'a}rdenas V{\'a}zquez},
  {Casal}, {Cifuentes}, {Claret}, {Colom{\'e}}, {Cort{\'e}s-Contreras},
  {Czesla}, {D{\'{\i}}ez-Alonso}, {Feiz}, {Fern{\'a}ndez}, {Ferro},
  {Fuhrmeister}, {Galad{\'{\i}}-Enr{\'{\i}}quez}, {Garcia-Piquer},
  {Garc{\'{\i}}a Vargas}, {Gesa}, {G{\'o}mez Galera}, {Gonz{\'a}lez-Peinado},
  {Gr{\"o}zinger}, {Grohnert}, {Gu{\`a}rdia}, {Guijarro}, {de Guindos},
  {Guti{\'e}rrez-Soto}, {Hagen}, {Hauschildt}, {Hedrosa}, {Helmling},
  {Hermelo}, {Hern{\'a}ndez Arab{\'{\i}}}, {Hern{\'a}ndez Casta{\~n}o},
  {Hern{\'a}ndez Hernando}, {Herrero}, {Huber}, {Huke}, {Johnson}, {de Juan},
  {Kim}, {Klein}, {Kl{\"u}ter}, {Klutsch}, {Lafarga}, {Lamp{\'o}n}, {Lara},
  {Laun}, {Lemke}, {Lenzen}, {L{\'o}pez del Fresno}, {L{\'o}pez-Gonz{\'a}lez},
  {L{\'o}pez-Puertas}, {L{\'o}pez Salas}, {Luque}, {Mag{\'a}n Madinabeitia},
  {Mall}, {Mandel}, {Marfil}, {Mar{\'{\i}}n Molina}, {Maroto Fern{\'a}ndez},
  {Mart{\'{\i}}n}, {Mart{\'{\i}}n-Ruiz}, {Marvin}, {Mirabet}, {Moya},
  {Moreno-Raya}, {Nagel}, {Naranjo}, {Nortmann}, {Ofir}, {Oreiro}, {Pall{\'e}},
  {Panduro}, {Pascual}, {Passegger}, {Pedraz}, {P{\'e}rez-Calpena}, {P{\'e}rez
  Medialdea}, {Perger}, {Perryman}, {Pluto}, {Rabaza}, {Ram{\'o}n}, {Rebolo},
  {Redondo}, {Reinhardt}, {Rhode}, {Rix}, {Rodler}, {Rodr{\'{\i}}guez},
  {Rodr{\'{\i}}guez Trinidad}, {Rohloff}, {Rosich}, {Sadegi},
  {S{\'a}nchez-Blanco}, {S{\'a}nchez Carrasco}, {S{\'a}nchez-L{\'o}pez},
  {Sanz-Forcada}, {Sarkis}, {Sarmiento}, {Sch{\"a}fer}, {Schiller},
  {Sch{\"o}fer}, {Schweitzer}, {Solano}, {Stahl}, {Strachan}, {Su{\'a}rez},
  {Tabernero}, {Tala}, {Tulloch}, {Veredas}, {Vico Linares}, {Vilardell},
  {Wagner}, {Winkler}, {Wolthoff}, {Xu}, {Yan}, \& {Zapatero
  Osorio}}]{trifonov2018}
{Trifonov}, T., {K{\"u}rster}, M., {Zechmeister}, M., {et~al.} 2018, \aap, 609,
  A117

\bibitem[{{Udalski} {et~al.}(2018){Udalski}, {Ryu}, {Sajadian}, {Gould},
  {Mr{\'o}z}, {Poleski}, {Szyma{\'n}ski}, {Skowron}, {Soszy{\'n}ski},
  {Koz{\l}owski}, {Pietrukowicz}, {Ulaczyk}, {Pawlak}, {Rybicki}, {Iwanek},
  {Albrow}, {Chung}, {Han}, {Hwang}, {Jung}, {Shin}, {Shvartzvald}, {Yee},
  {Zang}, {Zhu}, {Cha}, {Kim}, {Kim}, {Kim}, {Lee}, {Lee}, {Lee}, {Park},
  {Pogge}, {Bozza}, {Dominik}, {Helling}, {Hundertmark}, {J{\o}rgensen},
  {Longa-Pe{\~n}a}, {Lowry}, {Burgdorf}, {Campbell-White}, {Ciceri}, {Evans},
  {Figuera Jaimes}, {Fujii}, {Haikala}, {Henning}, {Hinse}, {Mancini},
  {Peixinho}, {Rahvar}, {Rabus}, {Skottfelt}, {Snodgrass}, {Southworth}, \&
  {von Essen}}]{udalski2018}
{Udalski}, A., {Ryu}, Y.~H., {Sajadian}, S., {et~al.} 2018, \actaa, 68, 1

\bibitem[{{Villanueva} {et~al.}(2018){Villanueva}, {Smith}, {Protopapa},
  {Faggi}, \& {Mandell}}]{villanueva2018}
{Villanueva}, G.~L., {Smith}, M.~D., {Protopapa}, S., {Faggi}, S., \&
  {Mandell}, A.~M. 2018, Journal of Quantitative Spectroscopy and Radiative
  Transfer, 217, 86

\bibitem[{{von Braun} {et~al.}(2014){von Braun}, {Boyajian}, {van Belle},
  {Kane}, {Jones}, {Farrington}, {Schaefer}, {Vargas}, {Scott}, {ten
  Brummelaar}, {Kephart}, {Gies}, {Ciardi}, {L{\'o}pez-Morales}, {Mazingue},
  {McAlister}, {Ridgway}, {Goldfinger}, {Turner}, \& {Sturmann}}]{vonbraun2014}
{von Braun}, K., {Boyajian}, T.~S., {van Belle}, G.~T., {et~al.} 2014, \mnras,
  438, 2413

\bibitem[{{Wenger} {et~al.}(2000){Wenger}, {Ochsenbein}, {Egret}, {Dubois},
  {Bonnarel}, {Borde}, {Genova}, {Jasniewicz}, {Lalo{\"e}}, {Lesteven}, \&
  {Monier}}]{wenger2000}
{Wenger}, M., {Ochsenbein}, F., {Egret}, D., {et~al.} 2000, Astronomy and
  Astrophysics Supplement Series, 143, 9

\bibitem[{{Wittenmyer} {et~al.}(2014){Wittenmyer}, {Tuomi}, {Butler}, {Jones},
  {Anglada-Escud{\'e}}, {Horner}, {Tinney}, {Marshall}, {Carter}, {Bailey},
  {Salter}, {O'Toole}, {Wright}, {Crane}, {Schectman}, {Arriagada}, {Thompson},
  {Minniti}, {Jenkins}, \& {Diaz}}]{wittenmyer2014}
{Wittenmyer}, R.~A., {Tuomi}, M., {Butler}, R.~P., {et~al.} 2014, \apj, 791,
  114

\bibitem[{{Wittenmyer} {et~al.}(2016){Wittenmyer}, {Butler}, {Tinney},
  {Horner}, {Carter}, {Wright}, {Jones}, {Bailey}, \&
  {O'Toole}}]{wittenmyer2016}
{Wittenmyer}, R.~A., {Butler}, R.~P., {Tinney}, C.~G., {et~al.} 2016, \apj,
  819, 28

\bibitem[{{Zhu} {et~al.}(2017){Zhu}, {Udalski}, {Calchi Novati}, {Chung},
  {Jung}, {Ryu}, {Shin}, {Gould}, {Lee}, {Albrow}, {Yee}, {Han}, {Hwang},
  {Cha}, {Kim}, {Kim}, {Kim}, {Kim}, {Lee}, {Park}, {Pogge}, {KMTNet
  Collaboration}, {Poleski}, {Mr{\'o}z}, {Pietrukowicz}, {Skowron},
  {Szyma{\'n}ski}, {KozLowski}, {Ulaczyk}, {Pawlak}, {OGLE Collaboration},
  {Beichman}, {Bryden}, {Carey}, {Fausnaugh}, {Gaudi}, {Henderson},
  {Shvartzvald}, {Wibking}, \& {Spitzer Team}}]{zhu2017}
{Zhu}, W., {Udalski}, A., {Calchi Novati}, S., {et~al.} 2017, \aj, 154, 210

\end{thebibliography}
\bibliographystyle{aasjournal}

\end{document}